\documentclass{article}
\usepackage[%
journal=XXX,    
lang=british,   
]{ems-journal}

\usepackage{bbm}
\usepackage{epsfig}
\usepackage[all]{xy}

\def\={\ =\ }
\def\dd{{\text{d}}}

\newcommand{\Tr}[1]{\:{\text{Tr}}#1}
\newcommand{\e}{{\,\text{e}\,}}
\newcommand{\mbf}[1]{{\boldsymbol {#1} }}

\def\ii{{\,\text{i}\,}}

\newcommand{\midwedge}{\text{\Large$\wedge$}}

\newcommand{\CA}{\mathcal{A}}    			
\def\alg{{\mathcal A}}

\def\hil{{\mathcal H}}
\def\bun{{\mathcal E}}

\newcommand{\CT}{{\mathcal{T}}}

\newcommand{\CB}{\mathcal{B}}

\newcommand{\CK}{\mathcal{K}}

\newcommand{\CM}{\mathcal{M}}

\newcommand{\CU}{\mathcal{U}}

\newcommand{\frg}{\mathfrak{g}}

\newcommand{\cald}{\mathcal{D}}

\newcommand{\ttA}{\mathtt{A}}

\newcommand{\bbr}{\mathbbm{R}}

\newcommand{\bbc}{\mathbbm{C}}

\newcommand{\bbT}{\mathbbm{T}}

\newcommand{\IZ}{\mathbbm{Z}}

\newcommand{\bbz}{{\mathbbm Z}}

\newcommand{\bbt}{{\mathbbm T}}

\newcommand{\longtwoheadrightarrow}{\,\relbar\joinrel\twoheadrightarrow\,}

\newcommand{\unit}{\mathbbm{1}}   			

\theoremstyle{plain}

\theoremstyle{definition}

\numberwithin{equation}{section}

\begin{document}

\title{Noncommutative Geometry of \\ Gravity, Strings and Fields: A Panoramic Overview}
\titlemark{Noncommutative Geometry of Gravity, Strings and Fields}



\emsauthor{1}{
	\givenname{Richard~J.}
	\surname{Szabo}
	\orcid{0000-0003-3821-4891}}{R.~J.~Szabo}

\Emsaffil{1}{
	\department{Department of Mathematics and Maxwell Institute for Mathematical Sciences}
	\organisation{Heriot-Watt University}
	\address{Riccarton}
	\zip{EH14 4AS}
	\city{Edinburgh}
	\country{United Kingdom}
	\affemail{R.J.Szabo@hw.ac.uk}}

\classification[46L87, 58B34, 46L85]{81T75}


\begin{abstract}
These are expanded lecture notes of a mini-course whose objectives were to introduce the basic concepts, constructions and techniques of noncommutative geometry, as well as their uses as a framework for modelling quantum spacetime. Key mathematical approaches presented include operator algebras such as $C^*$-algebras, K-theory, spectral geometry, quantum groups, and deformation quantization. Physical application areas considered include string theory, quantum field theory, and the Standard Model, as well as certain condensed matter systems. 
\end{abstract}

\maketitle

\tableofcontents


\section{Introduction}
\label{sec:Intro}

This contribution is based on a mini-course during the conference ``Applications of Noncommutative Geometry to Gauge Theories, Field Theories, and Quantum Spacetime” which was held at Centre International de Rencontres Math\'ematiques (CIRM) in Marseille from 7--11 April 2025. The lectures were delivered at the start of the conference, with the goal of  giving a broad introductory overview on noncommutative geometry and its applications in physics, as a means of induction into the more specialised mini-courses and talks that followed. This contribution is kept in much the same spirit, except that many details have been expanded on and key concepts elucidated more thoroughly. The material covered is for the most part standard, although some of the presentation and insights may be new to even readers with a solid background in the subject.

The goal of the lectures to some extent was to answer, through (some) theoretical development and (mostly) illustration via specific examples from physics, the following two questions:
\begin{enumerate}
\item What is noncommutative geometry? \label{Q1}

\item What is noncommutative physics?  \label{Q2}
\end{enumerate}
An attempt to Google these questions brings up responses that would make any expert cringe. The corresponding entries on {\sl Wikepedia} are especially inaccurate, trivial and useless, in desperate need of a complete rewrite. {\sl ChatGPT} outputs a slightly more palatable response, which however is still plagued with generic statements and unhelpfully vague summaries.

Let us start with the first question. Noncommutative geometry is a branch of mathematics that reformulates the foundations of classical geometry and topology by using purely algebraic tools that admit generalization to the theory of noncommutative algebras, thereby extending to `noncommutative spaces'. The mathematical framework was largely developed by Connes in the 1980s, who used it to provide a new approach to geometry and topology that connects with quantum theory. The generalisation avoids the distinction between continuous and discrete spaces, and allows treatment of singular spaces. It has had profound implications for the theory of foliations and groupoids, index theory, mathematical theories of quantization,  K-theory of group algebras, quantum groups, motivic number theory and the Riemann hypothesis, among many other problems in mathematics.

Noncommutative geometry has also become influential in various branches of physics, which brings us to the second question. It has proved especially useful for understanding the mathematical structure of quantum spacetime, a notion which permeates the quest for a suitable foundation of quantum gravity, where spacetime itself is expected to exhibit noncommutative features at very small (Planck scale) distances. In certain models, quantum field theories on noncommutative spaces, often referred to as `noncommutative field theories', are used to describe the interactions of particles at these scales. One notable example is string theory, a candidate theory of quantum gravity, which comes with an intrinsic minimal length scale and whose low energy limit is a field theory. In noncommutative field theory, field operators at separated points do not commute and modify the usual field equations, as well as the locality and behaviour of the interactions. This leads to novel quantum phenomena, as well as new types of symmetries.

Besides its motivation from the problem of quantising gravity and the related esoterics of string theory, noncommutative physics has also had diverse implications in studying various physical features at the quantum level, particularly in quantum mechanics, condensed matter physics and quantum field theory. Noncommutative geometry naturally describes the behaviour exhibited by some condensed matter systems, such as the quantum Hall effect and certain classes of topological insulators. The geometrisation of the Standard Model of fundamental interactions in particle physics culminated in the structure of a real spectral triple, which captures the metric aspects of noncommutative geometry. Hopf algebras presented an unexpected link between the noncommutative geometry of foliations and renormalization in perturbative quantum field theory. Noncommutative physics may also be relevant to quantum computing and quantum information theory.

In the following sections we provide an introduction to noncommutative physics by flushing out details surrounding some of these topics, as a panoramic overview of selected responses to the questions \eqref{Q1} and \eqref{Q2}.

\paragraph{Disclaimer.}

The field of noncommutative geometry is huge, as are its applications in physics. It would be impossible to review the entire area in simply a chapter contribution to a special volume. The choice of material covered in the following is largely personally biased to the authors' understanding and expertise in the area, in an attempt to avoid inaccurate or imprecise statements surrounding other works, rather than as an indication of relative importance or significance of topics. We do not even attempt to provide a very lengthy list of omitted topics here. Similarly, the literature on the subject is vast and lengthy, and we have mostly cited original articles on the topics covered. We apologise to those authors whose works we have not mentioned.

Early reviews of noncommutative field theories in similar spirits as the present contribution are found in~\cite{Douglas:2001ba,Szabo:2001kg}, while a more recent review is~\cite{Hersent:2022gry}, all with extensive bibliographies.

\paragraph{Outline.}

The outline by section of the content throughout the remainder of this contribution is as follows:
\begin{itemize}
\item[{\bf\underline{\S\ref{sec:spacetimequant}}}] We begin with an introduction to quantum spacetimes and their relevance in quantum gravity, describing models based on emergence and noncommutative geometry, along with a discussion of their phenomenology.

\item[{\bf\underline{\S\ref{sec:formalNCG}}}] We turn to mathematical aspects and review the basic ideas and constructions of noncommutative geometry, including Gel'fand-Naimark and Serre-Swan dualities, spectral triples, differential calculi from Dirac operators, and Morita equivalence, among others. We discuss the role of noncommutative geometry in high energy physics, including Connes' approach to the Standard Model coupled with gravity.

\item[{\bf\underline{\S\ref{sec:quantumonquantized}}}] We provide an early glimpse into noncommutative field theory, illustrated via a simple situation in quantum mechanics which demonstrates through elementary computations how such theories can arise as low energy effective field theories of real physical systems. 

\item[{\bf\underline{\S\ref{sec:strings}}}] We turn to aspects of string theory, and in particular the emergence of star-products of fields and noncommutative Yang-Mills theory. We discuss how the stringy symmetry of T-duality is `geometrised' in this framework, particularly in the setting of non-geometric backgrounds, through dual noncommutative Yang-Mills theories. A detailed prior knowledge of string theory is \emph{not} a prerequisite for understanding our presentation. 
 
\item[{\bf\underline{\S\ref{sec:defquant}}}] We look at two different approaches to deformation quantization and its physical applications. We start with the Kontsevich quantization formula and its relation to Seiberg-Witten maps, which realise the commutative/noncommutative duality of effective actions for strings and D-branes. We then pass to the Drinfel'd twist formalism and its relation to noncommutative gravity, illustrating the role of quantum groups as descriptions of quantum symmetries of physical models.

\item[{\bf\underline{\S\ref{sec:formalQFT}}}] We study some formal aspects of the perturbative expansion of noncommutative quantum field theory in the star-product formulation, focusing on their regularization and renormalization. We describe the notorious UV/IR mixing problem in some detail as well as some of its resolutions. 

\item[{\bf\underline{\S\ref{sec:matrixmodels}}}] We conclude  with a discussion of the intimate relationship between noncommutative field theories and matrix models, focusing on their descriptions as theories of emergent (noncommutative) spacetime and gravity, as well as the constructions of fuzzy field theories.
\end{itemize}

This article was written without any use of generative AI software.

\paragraph{Notation and Conventions.}

Unless otherwise explicitly stated, repeated indices are always implicitly summed over (whether upper/lower or not).

By an algebra we will always mean an associative algebra over the field of complex numbers $\bbc$, unless otherwise indicated. All differential graded algebras, as well as $L_\infty$-algebras, are $\mathbbm{N}_0$-graded.

\section{Spacetime Quantization}
\label{sec:spacetimequant}

Our tour through the applications of noncommutative geometry to physics will be, as disclaimed in \S\ref{sec:Intro}, personally biased towards the way in which geometry is deformed within certain models of quantum gravity. In this section we start with a general introduction to the main ideas and motivation behind these structures, before exploring them in more depth in later sections.

\subsection{Quantum Geometry from Quantum Gravity}

At the forefront of our understanding of theoretical high-energy physics rests the problem of unification of the two pillars of modern physics: general relativity and quantum mechanics. General relativity describes spacetime as a dynamical entity governed by Einstein's field equations, while quantum field theory treats spacetime as a fixed classical background. Classical general relativity
  breaks down at length scales of the order of the Planck scale $${ \ell_{\text P} \=\sqrt{\frac{\hbar\,G}{c^3}}~\sim~1.6\times
10^{-33}~\mbox{cm} }$$
where quantum effects become relevant. As a quantum field theory, general relativity with the Einstein-Hilbert action is non-renormalisable.
Quantum gravity is the so far elusive theory which attempts to incorporate gravity into a quantum theory.

There are many programmes for quantum gravity. Several approaches adopt the perspective that spacetime itself may not be fundamental, but is instead a coarse grained approximation of more primitive pre-geometric degrees of freedom. Then gravity is an \emph{emergent} phenomenon, arising through the interactions of more elementary constituents, and one should attempt to quantize the fundamental degrees of freedom from which gravity emerges. 

Emergent spacetime is a scenario permeating the celebrated Anti-de~Sitter/Conformal Field Theory (AdS/CFT) correspondence (see~\cite{Aharony:1999ti} for an early review), which is a concrete realization of a holographic principle for quantum gravity as well as the old idea that Yang-Mills theory in the limit of a large number of colours can be formulated as a string theory. In AdS/CFT holography, the bulk spacetime geometry and gravity emerge from the entanglement of quantum states on the boundary of spacetime,  through the duality between  string theory on anti-de~Sitter space and $\mathcal{N}=4$ supersymmetric Yang-Mills theory (a non-gravitational conformal field theory) on its boundary.

Another perspective common to many approaches, and the one taken in the present contribution, is through the notion of \emph{quantum geometry}, which applies the principles of quantum mechanics to spacetime itself. The basic premise behind this point of view is that, since gravity affects spacetime geometry according Einstein's theory, \emph{quantum} gravity should
\emph{quantize} spacetime, in some yet undetermined way. 

The simplest realisation of such a quantization would be to promote the coordinates $x^\mu$ of spacetime to noncommuting operators which obey commutation relations
\begin{align}\label{eq:MoyalWeylcomm}
{ \big[x^\mu\,,\,x^\nu\big]\=\ii\theta^{\mu\nu} } 
\end{align}
for a small constant antisymmetric deformation tensor $\theta^{\mu\nu}$ (of order $\ell^2_{\text P}$) that plays a role analogous to that of Planck's constant {$\hbar$} in phase space
quantization
$$\big[x^\mu\,,\,p_\nu\big]\=\ii\hbar\,\delta^\mu{}_\nu \ . $$ In particular, following the textbook derivation of the Heisenberg uncertainty principle in quantum mechanics, this  leads to spacetime uncertainty relations given by
$$
{ \Delta x^\mu\,\Delta
  x^\nu~\geqslant ~\mbox{$\frac12$}\, \big|\theta^{\mu\nu}\big| } \ ,
$$
which forbid the simultaneous measurement of spacetime coordinates. Thus the notion of a `point' is lost in quantized spacetime, and we may refer to its geometry as a ``pointless'' geometry (a term coined by von~Neumann to describe the geometry of quantum phase space). Modifications of this simple proposal, some of which are discussed below, have been relevant to many scenarios in quantum gravity phenomenology~\cite{Addazi:2021xuf,AlvesBatista:2023wqm}.

Following~\cite{Doplicher:1994tu}, one may construct a simple Gedanken experiment which illustrates that such noncommutativity of spacetime seems to be a model-independent feature of quantum gravity. Suppose one wishes probe physics at the Planck scale {$\ell_{\text P}$}. Then the Compton
wavelength of the probe must be smaller than ${ \ell_{\text P}}$. Using the de~Broglie relation, this implies that the probe must have a huge  mass ${m\geqslant \hbar/\ell_{\text
    P}\,c }$, and since it is concentrated in a tiny volume ${\ell_{\text P}^3}$ it forms
a {black hole}. The event horizon of the black hole, which has a Schwarzschild radius of order $ m$, then hides the measurement. Thus spacetime uncertainty principles are an inevitable feature of any model of Planck scale physics.

Fundamental uncertainties are in fact a common feature of many approaches to quantum gravity:
\begin{itemize}
\item Loop quantum gravity comes with minimal area and volume uncertainties, because geometric quantities like area and volume are quantized to operators~\cite{Ashtekar:2021kfp}.
\item {String theory} quantitatively predicts minimal length uncertainty relations as a feature evident in high-energy scattering amplitudes of strings, due to their finite intrinsic length scale $\ell_s$~\cite{Amati:1988tn}. We shall say a lot more about the role of noncommutative geometry in string theory in \S\ref{sec:strings}.
\item Some spin foam models naturally lead to effective noncommutative quantum field theories; we will look at a simple example in \S\S\ref{sub:3Dgravity} below.
\item Other  approaches to quantum gravity worth mentioning include 
group field theory~\cite{Freidel:2005qe}, causal dynamical triangulations~\cite{Loll:2019rdj},  causal set theory~\cite{Surya:2025knk}, and Ho\v{r}ava-Lifshitz gravity~\cite{Wang:2017brl}, among many others.
\end{itemize}

It is natural to suppose that \emph{quantum} spacetimes  come with \emph{quantum} symmetries. These are sometimes described as  deformations of classical Poincar\'e symmetry, often using the theory of quantum groups, among other mathematical techniques. The idea of quantum groups acting on quantum spaces  borrows techniques that have been largely developed in  the context of quantum integrable systems and the search for solutions to the quantum Yang-Baxter equation, which are based on Hopf algebras and Drinfel'd twist quantization. We will give a more detailed introduction in~\S\S\ref{sec:Drinfeldtwist}.
 
The mathematical framework of \emph{noncommutative geometry} was developed in the 1980s independently by Connes~\cite{Connes85}, Woronowicz~\cite{Woronowicz:1987wr}, and Majid~\cite{Majid:1988we}, among others. In physics it has blossomed into a concrete framework for quantum geometry aimed at understanding certain regimes of quantum gravity; as we will see in \S\S\ref{sub:twistedreduced}, the framework naturally encompasses the notions of spacetime and gravity as emergent phenomena. Noncommutative geometry generalizes the duality between spaces and algebras, allowing for the description of quantum spacetimes that lack a classical point set; this duality is spelled out in detail in \S\S\ref{sub:spectral_description}. The theory also provides useful and convenient settings for the structures of   more solidly grounded effective theories in condensed matter and high energy physics, some of which are amenable to current experimental comparison. It is in this spirit that the present overview is focused. 

For the remainder of this introductory section, we highlight three concrete examples of quantized spacetime to illustrate some of the above features more quantitatively.

\subsection{$\kappa$-Minkowski Space}
\label{sub:kappaMink}

One of the most extensively studied models of a quantum geometry is $\kappa$-Minkowski space, which was originally introduced as a module for the  $\kappa$-Poincar\'e quantum group in~\cite{Lukierski:1992dt,Majid:1994cy}. Compared to the `canonical' spacetime commutation relations \eqref{eq:MoyalWeylcomm}, which involve the standard canonically conjugate position and momentum operators, the full quantum phase space is now modified to
\begin{align*}
\big[x^\mu\,,\,x^\nu\big]&\=\frac\ii{{\kappa}} \,\big(x^\mu\,\xi^\nu-
x^\nu\,\xi^\mu\big) \ , \\[4pt]
\big[x^\mu\,,\,p_\nu\big]&\=\ii\hbar\,\delta^\mu{}_\nu+
\frac\ii{{\kappa}}\,
\big(p^\mu\,\xi_\nu+p_\nu\,\xi^\mu\big) \ , \\[4pt]
\big[p_\mu\,,\,p_\nu\big]&\=0 \ ,
\end{align*}
where the light-like vector $\xi^\mu$ has components {$\xi^0=1$} and {$\xi^i=0$}, while the deformation parameter {$\kappa$} is a mass
scale which we identify with $\hbar/\ell_{\text{P}}$.

The deformation of the position-momentum commutators ensures that the brackets satisfy the Jacobi identity, and moreover that they can be inverted to give a symplectic structure on  phase space. The position commutators break classical Poincar\'e invariance to a $\kappa$-Poincar\'e symmetry. As translation invariance is lost, so is standard momentum conservation, which is now deformed to a conservation law imposed by deforming the sum of vectors in momentum space (via the Baker-Campbell-Hausdorff formula applied to products of plane wave operators). In particular, this leads to modified dispersion relations of the form
\begin{align}\label{eq:kappadispersion}
E^2\=\mbf p^2\,c^2-m^2\,c^4\,\Big(1-\frac{\xi\cdot p}
{\hbar\,{\kappa}}\Big)^2 \ ,
\end{align}
for a relativistic particle of mass $m$ with momentum $p=(E,\mbf p)$.

This model of quantum spacetime has been discussed as a concrete realisation of doubly special relativity and the relative locality principle, in which the speed of a photon depends on its energy {$E$} through the combination
{$\ell_{\text{P}}\,E=\hbar\,E/\kappa$}.   The deformed dispersion relations can be matched phenomenologically with measurements of astrophysical gamma-ray bursts from the Gamma-ray Large Area Space Telescope (GLAST)~\cite{Amelino-Camelia:2000cpa,Kowalski-Glikman:2002iba,Magueijo:2001cr}.

\subsection{Three-Dimensional Quantum Gravity}
\label{sub:3Dgravity}

It was realised long ago that quantum gravity in three dimensions provides a concrete model of quantum spacetime, going back to the works~\cite{tHooft:1996ziz,Matschull:1997du}. There are no gravitons in three spacetime dimensions, so three-dimensional general relativity is an exactly solvable system. Propagating degrees of freedom are introduced by coupling gravity to matter. 

In particular, integrating out the gravitational degrees of freedom in a certain spin foam model of three-dimensional gravity  coupled to spinless matter leads to an effective scalar field theory on
an  $SO(1,2)$ Lie algebra noncommutative spacetime~\cite{Freidel:2005me}. The quantum phase space commutators are given by
\begin{align*}
\big[x^\mu\,,\,x^\nu\big]&\=\ii{\ell_{\text P}}\,\epsilon^{\mu\nu\lambda}\,
x^\lambda \ , \\[4pt]
\big[x_\mu\,,\,p_\nu\big]&\=\ii\sqrt{\hbar^2-{\ell_{\text P}^2}
\,p^2}~\delta_{\mu\nu}-
\ii{\ell_{\text P}}\,\epsilon_{\mu\nu\lambda}\,p_\lambda \ , \\[4pt]
\big[p_\mu\,,\,p_\nu\big]&\=0 \ ,
\end{align*}
which again satisfy the Jacobi identity and can be inverted to a symplectic structure. 

Again the breaking of translation invariance implies a deformed momentum conservation law, leading in particular to the dispersion relations
\begin{align}\label{eq:so21dispersion}
E^2\=\mbf p^2\,c^2-\Big(\frac{\sinh({\ell_{\text P}}\,m\,c^2/\hbar)}
{{\ell_{\text P}/\hbar}}\Big)^2 \ .
\end{align}
By comparing \eqref{eq:so21dispersion} with \eqref{eq:kappadispersion}, this model has thus been used to suggest that doubly  special relativity arises in the low energy limit of quantum gravity.

\subsection{Snyder's Spacetime}

For the sake of historical completeness, let us take as a final example the very first proposal of quantized spacetime from the 1940s, due to Snyder~\cite{Snyder:1946qz,Snyder:1947nq,Yang:1947ud}. In contrast to the models of \S\S\ref{sub:kappaMink} and \S\S\ref{sub:3Dgravity} above, Snyder's spacetime retains classical Poincar\'e invariance. The original motivation was to use noncommutative geometry to tame the ultraviolet divergences of quantum field theory, as an alternative to cutoff, lattice or dimensional regularisation schemes which break Lorentz invariance. With the success of renormalization theory in the early 1950s, Snyder's program never matured. In fact, as we will discuss later on, spacetime noncommutativity does not necessarily cure problems with ultraviolet divergences.

Snyder's algebra is based on the commutation relations
\begin{align*}
\big[x^\mu\,,\,x^\nu\big] & \= \ii\ell^2_\text{P} \, M^{\mu\nu} \ . 
\end{align*}
Demanding that the symmetries of this spacetime geometry are described by an undeformed Poincar\'e symmetry implies that both the Lorentz generators $M_{\mu\nu}=-M_{\nu\mu}$ and the translation generators $p_\mu$ satisfy the standard commutation relations
\begin{align*}
\big[M_{\mu\nu}\,,\,M_{\lambda\rho}\big] & \= \ii\big(\eta_{\nu\lambda}\,M_{\mu\rho} - \eta_{\rho\lambda}\,M_{\nu\rho} - \eta_{\nu\rho}\,M_{\rho\lambda} + \eta_{\mu\rho}\,M_{\nu\lambda}\big) \ , \\[4pt]
\big[p_\mu\,,\,p_\nu\big]&\=0 \ .
\end{align*}
We further assume the noncommutative coordinates and the momenta transform as undeformed vectors under the Lorentz algebra, which implies the commutators
\begin{align*}
\big[M^{\mu\nu}\,,\,x^\lambda\big] & \= \ii\big(\eta^{\nu\lambda}\,x^\mu - \eta^{\mu\lambda}\,x^\nu\big) \ , \\[4pt]
\big[M_{\mu\nu}\,,\,p_\lambda\big] & \= \ii\big(\eta_{\nu\lambda}\,p_\mu - \eta_{\mu\lambda}\,p_\nu\big) \ .
\end{align*}
The quantity $p^2=\eta^{\mu\nu}\,p_\mu\,p_\nu$ is then a Lorentz invariant.

Interest in the Snyder spacetime has been rekindled in more recent times due its relations with scenarios of doubly special relativity~\cite{Kowalski-Glikman:2002eyl,Guo:2008qp}, loop quantum gravity~\cite{Livine:2004xy} and two-time physics~\cite{Romero:2004er}.

\section{Mathematical Basis of Quantum Geometry}
\label{sec:formalNCG}

In this section we give a brief introduction to some of the mathematical concepts of noncommutative geometry, which are crucially influenced by quantum
physics, as well as some of its basic applications in particle physics. Later on we will use this formalism to accurately describe the facets of quantum geometry as a suitable foundation of quantum gravity. In order to get across the main ideas and conceptual aspects, we shall usually gloss over many technical details, and sometimes be slightly imprecise in our mathematical statements. For more in depth and accurate treatments of the subject, the reader is invited to consult the classic texts~\cite{Connes:1994yd,Landi:1997sh,Madore:2000aq,Gracia-Bondia:2001upu,ConnesMarcolli2008,BeggsMajid2020,vanSuijlekom:2024jvw}.

\subsection{The Spectral Description}
\label{sub:spectral_description}

Classical Riemmanian geometry is based on introducing the notion of distance on arbitrary manifolds $M$ through a metric tensor $g$ on $M$ with positive signature. The geodesic distance $d(x,y)$ between any two points $x,y\in M$ may then be computed from the formula
\begin{align} \label{eq:geodesic}
d(x,y)\=\inf_{\gamma:x\to y} \ \int_{\gamma}\,\dd s \ ,
\end{align}
where the infemum is taken over all paths $\gamma\subset M$ between $x$ and $y$, and the Riemannian line element $\dd s$ is given in local coordinates by
\begin{align} \label{eq:linelement}
\dd s^2\=g_{ij}\,\dd x^i\,\dd x^j \ .
\end{align}

The starting point of noncommutative geometry is to translate this classical description into a purely algebraic language, akin to what happens in the passage from classical to quantum physics. This is possible thanks to a duality between topological spaces and algebras, known as the {\it Gel'fand-Naimark Theorem}: There is a one-to-one correspondence between compact \emph{Hausdorff} spaces $M$ and \emph{commutative} $C^*$-algebras $\CA$. By dropping the prefix `$C^*$-', as we shall usually do for brevity, the correspondence extends more generally to non-compact spaces and even to non-Hausdorff spaces as well as to schemes (where only the underlying ring structure of $\CA$ matters). Let us describe the basis of this duality.

Starting from a commutative algebra $\CA$ over $\bbc$, a topological space may recovered from its \emph{spectrum} 
\begin{align*}
M\=\text{Spec}(\CA) \ ,
\end{align*}
which is the set of equivalence classes of non-trivial irreducible representations of $\CA$ on a Hilbert space $\hil$. A topology on $M$ is introduced by considering the prime ideals of $\CA$, where an ideal is prime if it is the kernel of a non-trivial irreducible representation. The set of prime ideals $\text{Prim}(\CA)$ is a topological space with the Jacobson topology, and for a commutative algebra $\CA$ there is a natural bijection $$\text{Spec}(\CA) \ \simeq \ \text{Prim}(\CA) \ . $$

The set of all points $x\in M$ is then recovered from the set of all characters  of the algebra $\CA$, i.e. the linear multiplicative maps $\chi:\CA\longrightarrow\bbc$. This identifies 
\begin{align*}
\CA\=C(M)
\end{align*}
with the commutative algebra of continuous complex-valued functions on $M$ with pointwise multiplication $f\cdot g$ for $f,g:M\longrightarrow\bbc$; when $M$ is compact this is a $C^*$-algebra with the $L^\infty$ norm $$\|f\|\=\|f\|_\infty \ := \ \sup_{x\in M} \, |f(x)| \ . $$ The value of a character on $f\in\CA=C(M)$ corresponds to the evaluation
\begin{align*}
\chi_x(f) \= f(x) \ .
\end{align*}
Alternatively, the points of $M$ can be recovered from the prime ideals of $\CA=C(M)$ which are all of the form
\begin{align}\nonumber
I_x \= \ker(\chi_x)\=\{f\in C(M) \ | \ f(x)\=0\} \ ,
\end{align}
the maximal ideal of functions vanishing at $x\in M$. 

The Gel'fand-Naimark duality can be extended to smooth manifolds $M$ by considering the dense Fr\'echet subalgebra $C^\infty(M)\subset C(M)$ of smooth complex-valued functions. Moreover, it is a functorial equivalence of categories. This means that every commutative algebra can be realised as the functions on a manifold, and  every algebra homomorphism is induced by a unique smooth map of manifolds.

How does one now introduce a Riemannian metric $g$ on $M$? Starting from a Riemannian manifold $(M,g)$ with a spin structure, we can represent the algebra $\CA$ on the Hilbert space $$\hil\=L^2(M,S)$$ of square-integrable sections of the complex spin bundle  $S\longrightarrow M$, i.e. spinors on $M$, by pointwise multiplication. The curved space Dirac operator $$D\=\nabla^S\=\ii\gamma^i\,\nabla_i$$ acts on spinors, where $\nabla$ is the spin connection and the curved space gamma-matrices $\gamma^i(x)$ obey the anticommutation relations $$\{\gamma^i,\gamma^j\}\=2\,g^{ij} \ . $$
We may then recover the Riemannian line element \eqref{eq:linelement} formally from the Dirac propagator as
\begin{align}\nonumber
\dd s^{-1}\=D:\hil \ \longrightarrow \ \hil \ .
\end{align}
Self-adjointness of $D$ guarantees that $\dd s^2\geqslant0$.

The geodesic distance formula \eqref{eq:geodesic}  can now be rewritten in a purely algebraic fashion as
$$
d(x,y)\=\sup_{f\in\CA}\,\big\{|\chi_x(f)-\chi_y(f)| \ \big| \ \|[D,f]\|\leqslant 1\big\} \ .
$$
The intuition behind this formula is that one can always find a function $f:M\longrightarrow\bbc$ with a finite difference of its values on two points $x,y\in M$ such that $|f(x)-f(y)|=d(x,y)$ (by rescaling $f$ if necessary). The key issue is a canonical choice of such a function. This is achieved by looking at all functions whose gradients on $M$ are bounded from above by~$1$, and then selecting the one which maximizes the distance $|f(x)-f(y)|$ between the two points in $\bbc$. The Dirac operator $D$ supplies a suitable gradient.

This spectral description of Riemannian geometry has far reaching generalisations. By abstracting the properties of the Dirac operator $D$ as a first-order differential operator, one can introduce a notion of `generalised' Dirac operator for an arbitrary associative algebra $\CA$, commutative or noncommutative, represented faithfully on a Hilbert space $\hil$ (which is always possible): A generalised Dirac operator is a self-adjoint operator $D:\hil\longrightarrow\hil$ with compact resolvent such that the commutator $[D,a]$ is bounded for all $a\in\CA$. The basic data of  spectral geometry is thus encoded in a \emph{spectral triple}
$$\CT\=(\CA,\hil,D) \ , $$
consisting of an algebra $\CA$ that captures the topological data, a representation $\hil$ of $\CA$, and a generalised Dirac operator $D:\hil\longrightarrow\hil$ which encodes the metric data. 

When supplemented with  additional data such as a real structure $J:\hil\longrightarrow\hil$~\cite{Connes:1995tu}, Connes' Reconstruction Theorem asserts a one-to-one correspondence between compact spin manifolds and spectral triples where $\CA$ is a commutative algebra~\cite{Connes:1996gi,Connes:2008vs}. When $\hil$ is the Hilbert space of spinors on a spin manifold, $J$ is the charge conjugation operator. This repackaging of data is valuable for describing physical systems geometrically, as it provides a framework for constructing a geometry tailored to a given physical problem, rather than attempting to fit the physics into the restrictive framework of classical Riemannian geometry. This will be particularly relevant to our later discussions of stringy deformations of geometry in \S\ref{sec:strings}.
 
When the algebra $\CA$ is noncommutative, the Gel'fand-Naimark duality is lost: there is generally no longer any notion of a `point'. While the spectrum of $\CA$ still makes sense, and can be regarded as a suitable dual `space' or `scheme', there are generally more non-trivial irreducible representations of $\CA$ than there are prime ideals. Since equivalent representations have the same kernel, there is a surjection $$\text{Spec}(\CA) \ \longtwoheadrightarrow \ \text{Prim}(\CA)$$ under which the preimages of open sets in the Jacobson topology define a topology on the spectrum. Thus the `coordinate' algebra $\CA$ still provides the required topological information. 

In this purely algebraic framework, it is even possible to introduce a suitable noncommutative differential calculus~$\Omega_D^\bullet\CA$. This is the differential graded algebra over $\Omega_D^0\CA=\CA$, on which the differential is given by
\begin{align}\nonumber
\dd_D(a)\=[D,a] \ ,
\end{align}
for $a\in\CA$. Then the first order differential calculus of the spectral triple $\CT=(\CA,\hil,D)$ is given by the $\CA$-module of noncommutative one-forms
\begin{align}\nonumber
\Omega_D^1\CA\= \text{Span}_\CA\big\{[D,a] \ \big| \ a\in\CA\big\} \ .
\end{align}
Higher orders require quotients by ideals of `junk forms' in order to ensure that $\dd_D$ acts as a differential.

The representations of a generic $\bbc$-algebra $\CA$ also capture the coarsest `topological invariants' of $\CA$ in the absence of any further data. The $\CA$-modules together with $\CA$-linear maps form an abelian category, and the Grothendieck group of this category is called the K-theory group $K(\CA)$ of the algebra $\CA$. Concretely, $K(\CA)$ is the Grothendieck completion of the semi-group of stable isomorphism classes of $\CA$-modules with respect to direct sum of representations.

\subsection{Gauge Symmetries and Morita Equivalence}
\label{sub:Morita}

The passage to noncommutative algebras opens up many rich equivalences between spaces that have no classical analogues. These equivalences in noncommutative geometry can be used to naturally explain duality symmetries of physical theories that cannot be regarded as symmetries in any sense of classical geometry. 

To explain the idea in a form relevant to physical applications, let us consider the symmetries of a unital noncommutative $\bbc$-algebra $\CA$ endowed with an antilinear $*$-involution ${}^\dag:\CA\longrightarrow\CA$. From this data we may introduce the gauge group of $\CA$ as the group of unitary elements of the algebra
\begin{align} \nonumber
\mathcal{U}(\CA)\=\{u\in\CA \ | \ u^\dag\,u\=u\,u^\dag\=\unit\} \ .
\end{align}
The gauge group parametrizes gauge transformations
$$g_u:\CA \ \longrightarrow \ \CA \quad , \quad g_u(a)\=u\,a\,u^\dag$$
which generate the group of inner automorphisms $$\text{Inn}(\CA) \= \text{Ad}\big(\mathcal{U}(\CA)\big)$$ of the algebra $\CA$. This defines the automorphism group as the extension of the inner automorphisms by the group $\text{Out}(\CA)$ of outer automorphisms through the semi-direct product of infinite-dimensional Lie groups $$\text{Aut}(\CA)\=\text{Inn}(\CA)\rtimes\text{Out}(\CA) \ . $$ 

For any commutative algebra $\CA$, there are no non-trivial inner automorphisms, so $\text{Inn}(\CA)=\{\text{id}\} $ and $\text{Aut}(\CA)=\text{Out}(\CA)$. For the infinite-dimensional algebra $$\CA\=C^\infty(M)$$ of smooth functions on a manifold $M$,  with $*$-involution given by complex conjugation $$f^\dag(x) \ := \ \overline{f(x)} \ , $$ the automorphisms  $$\text{Aut}(\CA)\=\text{Diff}(M)$$ form the group of diffeomorphisms of $M$, acting on functions by pullback $$f \ \longmapsto \  f\circ\varphi^{-1}$$ of a function $f:M\longrightarrow \bbc$ by a diffeomorphism $\varphi:M\longrightarrow M$. 

Consider now the  mildly noncommutative  algebra
$$\CA_N\=C^\infty\big(M,\text{Mat}(N)\big) \ \simeq \ C^\infty(M)\otimes \text{Mat}(N)  $$
of smooth functions on $M$ valued in the finite-dimensional algebra $\text{Mat}(N)$ of $N\times N$ complex matrices, with the usual Hermitian conjugation as $*$-involution. The gauge group of the algebra $\text{Mat}(N)$ is the $N\times N$ unitary group
\begin{align}\nonumber
\mathcal{U}\big(\text{Mat}(N)\big) \= U(N) \ ,
\end{align}
whereas $\text{Mat}(N)$ has no non-trivial outer automorphisms: $\text{Out}(\text{Mat}(N))=\{\text{id}\}$.
Hence in this case
\begin{align*}
\text{Inn}(\CA_N)&\=\text{Ad}\big(C^\infty(M,U(N))\big) \ , \\[4pt] \text{Out}(\CA_N)&\=\text{Diff}(M) \ .
\end{align*}
The difference here from the commutative case $\CA=C^\infty(M)$  lies only in the fact that $\CA_N$ is the algebra of fields of some theory with an internal gauge symmetry, represented by the non-trivial inner automorphisms. But the spacetime symmetries represented by the outer automorphisms are unchanged, and clearly we would still like to identify $\CA_N$ as dual to the original space $M$ that we started with.

This intuition can indeed be made precise by noting that the matrix algebra $\text{Mat}(N)$ has only one non-trivial irreducible representation (as an associative algebra), namely its fundamental representation acting on the finite-dimensional Hilbert space $\hil_N=\bbc^N$. It follows that the algebras $\CA=C^\infty(M)$ and $\CA_N$ are both dual to the same underlying topological space: $$\text{Spec}(\CA_N) \ \simeq \ \text{Spec}(\CA)\=M \ . $$ We capture this duality by saying that  $\CA$ and $\CA_N$ are {\it Morita equivalent}. In this example, a stronger statement is in fact true: $\CA$ and $\CA_N$ are \emph{stably isomorphic}.

The formal notion of Morita equivalence between two algebras $\CA$ and $\CB$ is far more general and goes beyond stable isomorphism as well as isomorphic spectra. There are several versions: the weakest version states that  their abelian categories of modules are functorially equivalent, while a stronger  version states that $\CB$ is Morita equivalent to $\CA$ if $$\CB \ \simeq \ \text{End}_\CA(\mathscr{E})$$ can be identified with the algebra of $\CA$-linear maps of some $\CA$-module $\mathscr{E}$. But the basic consequence of all versions is the same: Morita equivalent algebras have equivalent representation theories. In particular, their K-theory groups are the same: $$K(\CA) \ \simeq \  K(\CB) \ . $$

Morita equivalence becomes particularly relevant in the context of field theories with internal symmetries, particularly gauge theories. This is thanks to a duality between complex vector bundles $E\longrightarrow M$ over a space $M$ and modules over the commutative algebra of functions $\CA=C(M)$: the sections of $E$ form an $\CA$-module with respect to pointwise multiplication of a section $\sigma:M\longrightarrow E$ with a function  $f:M\longrightarrow\bbc$. The {\it Serre-Swan Theorem} provides a one-to-one correspondence between finitely-generated projective modules over a commutative algebra $\CA$ and vector bundles over $M=\text{Spec}(\CA)$. This extends to an equivalence of categories under the global section functor which takes a vector bundle $E\longrightarrow M$ to its $\CA$-module of sections, and also to the smooth setting by the $C^\infty$-version of the Serre-Swan Theorem~\cite{Nestruev2020}. In particular, the K-theory group of $\CA=C(M)$ is naturally isomorphic to the K-theory group $K(M)$ of vector bundles over $M$, whereby the direct sum of $\CA$-modules corresponds to the Whitney sum of vector bundles.

\subsection{The Noncommutative Standard Model}
\label{sub:StandardModel}

Following our motivating example from \S\S\ref{sub:Morita}, the noncommutative geometries that capture the internal symmetries of gauge theories are based on finite-dimensional algebras. If $\CA_{\text F}$ is any finite-dimensional algebra over $\bbc$, then it is isomorphic to a finite direct sum of full matrix algebras, and there is a canonical isomorphism $$\text{Spec}(\CA_\text{F}) \ \simeq \ \text{Prim}(\CA_\text{F})$$ that endows the spectrum with the discrete topology. If $\CA$ is a commutative algebra, then $\CA\otimes\CA_\text{F}$ is said to be an {almost commutative} algebra, and noncommutative geometry based on  almost commutative algebras is called  {almost commutative geometry}. 

For an almost commutative geometry based on $$C(M)\otimes\CA_{\text{F}} \ , $$ the spacetime $M$ is augmented by a discrete `internal' space  $$F\=\text{Spec}(\CA_{\text{F}}) \ , $$ whose points correspond to the simple direct summands of $\CA_\text{F}$. The summands couple through the introduction of gauge fields based on noncommutative differential one-forms in $\Omega_D^1\CA_\text{F}$. 

The simplest example is furnished by the two-dimensional algebra $$\CA_{\text{F}}\=\bbc\,\oplus\,\bbc \ , $$ with gauge group  $$\mathcal{U}(\CA_\text{F})\=U(1)\times U(1)$$ corresponding to the Higgs sector of a spontaneously broken $U(2)$  symmetry. In this case $$M\times F\= M\sqcup M$$ is a two-sheeted spacetime. The Standard Model on a two-sheeted spacetime is the celebrated Connes-Lott model~\cite{Connes:1990qp}, in which the Higgs field naturally finds a geometric interpretation as  the component of the noncommutative gauge potential in the discrete direction.

That the internal spaces of  theories with multiple gauge sectors are connected by Higgs fields in this way has led to a variety of applications of almost commutative geometry to particle physics. The state-of-the-art provides a geometrical interpretation of the full Standard Model through the  finite-dimensional real algebra 
\begin{align}\notag
\CA_{\text{F}}\=\bbc\,\oplus\,\mathbb{H}_{\text R}\,\oplus\,\mathbb{H}_{\text L}\,\oplus\, \text{Mat}(3) \ ,
\end{align}
where $\mathbb{H}_{\text R}$ and $\mathbb{H}_{\text L}$ are two (right and left) copies of the algebra of quaternions. From the isomorphism of complex algebras $$\CA_\text{F} \ \simeq \ \bbc\,\oplus\,\text{Mat}(2)\,\oplus\,\text{Mat}(3)$$ it follows that its gauge group is $$\mathcal{U}(\CA_\text{F})\=U(1)\times U(2)\times U(3) \ . $$

The space $F=\text{Spec}(\CA_{\text{F}})$ in this  example has classical dimension $0$ but KO-dimension $6$ (mod~$8$), where `KO' stands for the K-theory of \emph{real} representations (vector bundles). On the other hand, the classical and KO-dimension of a smooth manifold $M$ coincide. In particular, when $M$ is four-dimensional, $M\times F$ is a noncommutative Kaluza-Klein compactification of a spacetime of KO-dimension  $4+6=10$~\cite{Connes:2006qv,Barrett:2006qq,Chamseddine:2006ep}. 
This  dimension is of course tantalizing as it coincides with the critical dimension of superstring theory, whose compactifications are also believed to provide realistic models of particle physics in suitable scenarios; we will begin discussing the noncommutative geometry of string theory in \S\ref{sec:strings}. 

In a similar vein, the spectral approach can also be used couple gravity  to the Standard Model through the spectral action~\cite{Connes:1996gi,Chamseddine:1996zu,Chamseddine:1996rw} which incorporates fluctuations of the Dirac operator $$D\longmapsto u\,D\,u^\dag \ , \quad u \ \in \ \CU(\CA) \ \subset \ U(\hil)$$ by gauge transformations; a more elaborate account can be found in the contribution of Ali Chamseddine to this volume~\cite{Chamseddine:2025wgr}. Extensions of spectral geometry in this context to nonassociative coordinate algebras $\CA$ of spectral triples are discussed in~\cite{Farnsworth:2013nza,Boyle:2014wba}; we shall also encounter physical models exhibiting nonassociative geometry starting in \S\ref{sec:quantumonquantized}.

\section{Quantum Theory on Quantized Spacetime}
\label{sec:quantumonquantized}

In this section we  provide a brief general overview of some of the physical features of field theory on quantized spacetimes, aspects of which will be discussed in more detail in subsequent sections. Then we will turn to some elementary concrete physical examples of quantum geometry, which arise in the quantum mechanics of charged particles in external fields; a more detailed account of these systems and their noncommutative geometries can be found in~\cite{Szabo:2004ic,Szabo:2019hhg}. This will lay the groundwork for some of the more elaborate models of quantum gravity that we consider later on, as well as introduce the general formalism for handling noncommutative field theories.

\subsection{Noncommutative Quantum Field Theory: A Prelude}

Quantum field theory as we currently understand it works well at least down to the scale probed by the Large Hadron Collider (LHC)
  $$\ell_{\text{LHC}} \ \sim \ 2\times10^{-18}~\text{cm} \ .$$
What happens to our description of physics below this scale? {\it Noncommutative quantum field theory} has been suggested as a framework which may be relevant at scales in
  between the Planck length ${ \ell_{\text P} }$ and ${ \ell_{\text{LHC}} }$. As such, it has several appealing features. Noncommutativity provides an alternative to supersymmetry and string theory, which by relating field theory to gravity suggests a possibly easier approach to quantizing general relativity. It also yields a concrete avenue for exploring physics beyond the Standard Model, and arises naturally as effective theories of standard physics in strong external fields; we will discuss  one notable example of the latter in this section.
  
Noncommutative field theories also exhibit a variety of new phenomena, which enable us to explore foundational issues in quantum field theory and compare predictions of noncommutativity with experimental results. An incomplete list of novel effects is as follows:
\begin{itemize}

\item The theories exhibit a controlled form of Lorentz violation involving localized field
  configurations within a fixed observer inertial frame. This can be compared with Lorentz-violating scenarios of present day physics which are well-constrained experimentally \cite{Bolmont:2022yad}. Lorentz invariance is instead manifested as a `twisted' symmetry, which we will say more about in \S\S\ref{sub:twistedsymmetries}.
  
  \item The theories exhibit violations of causality: there is no longer a sharply localized light-cone, but rather regions described as ``light-wedges''~\cite{Grosse:2007vr,Grosse:2008dk}.
  
  \item  The ultraviolet/infrared (UV/IR) mixing problem~\cite{Minwalla:1999px} renders most noncommutative field theories non-renormalizable. In particular, contrary to historical lore, spacetime coarse-graining does not generally tame the ultraviolet  divergences of quantum field theory. We shall say more about UV/IR mixing as well as some of its  cures in \S\ref{sec:formalQFT}.
  
  \item Noncommutativity can lead to new interactions as well as interactions which would otherwise be forbidden by Lorentz invariance, causality or unitarity.

\end{itemize}
These novel features have all led to a flurry of investigations into potential new phenomenology and their verifications in experiment, including quantum mechanics, model building in particle physics, as well as the Standard Model; see e.g.~\cite{Chaichian:2000si,Terashima:2000xq,Carroll:2001ws,Calmet:2001na,Amelino-Camelia:2001hzq,Chaichian:2001mu,Chaichian:2001py,Balachandran:2005eb,Helling:2007zv,Chaichian:2009uw} for an incomplete list of early works on this topic, and~\cite{Szabo:2009tn} for an early review.

\subsection{Quantum Mechanics in Magnetic Fields}
\label{sub:magneticcharge}

A simple and natural example of a noncommutative geometry arises in the quantum mechanics of an electrically charged particle, of charge $e$ and mass $m$, in three-dimensional space $\bbr^3$ in the presence of an external magnetic field. Let us first consider the case that the particle motion is confined to a plane $\bbr^2\subset\bbr^3$ with coordinates $\mbf x = (x^1,x^2)$ and subject to a perpendicularly applied constant magnetic field of strength $B$; this is known as the Landau problem. The Lagrangian of the particle is then
$$
\displaystyle \mathcal{L}_m\=\frac m2\,\dot{\mbf x}{}^2-\frac ec\,\dot{\mbf
  x}\cdot\mbf A -V(\mbf x) \ , $$ 
where an overdot indicates time derivative, $V(\mbf x)$ is an external potential representing possible impurities in the medium, and $\mbf A=(A_1,A_2)$ is the magnetic vector potential which in a symmetric gauge reads as  
  $$ \displaystyle A_i\=-\frac B2\, \epsilon_{ij}\, x^j \ .
$$
Here $\epsilon_{ij}$ is the two-dimensional Levi-Civita symbol with $\epsilon_{12}=+1$.

When $V(\mbf x)=0$, the quantum dynamics of this system can be mapped onto that of a simple harmonic oscillator, whose energy levels are known as Landau levels. The strong field limit {$e\,B\gg m$} projects onto the lowest Landau level, which is equivalent to taking the formal $m\to0$ limit of the Lagrangian to get
\begin{align}\label{eq:calL0}
\mathcal{L}_0\=-\frac{e\,B}{2c}\,\dot x^i\,\epsilon_{ij}\, x^j - V(\mbf x) \ .
\end{align}
This Lagrangian is of first order in time derivatives, so it leads to a degenerate phase space in which the canonically conjugate momenta coincide with the position coordinates. 

Canonical quantization of the Poisson brackets in the lowest Landau level thus gives {\it noncommuting coordinates} with the commutation relations
$$
[x^i,x^j]\=\ii\theta^{ij} \ := \ \frac{\ii\hbar\,c}{e\,B}\, \epsilon^{ij} \ .
$$
In a Schr\"odinger polarisation, these relations are satisfied by differential operators 
\begin{align}\label{eq:Schrx}
\widehat x{}^i \= x^i-\frac{\ii}{2}\, \theta^{ij}\,\frac\partial{\partial x^j} \ .
\end{align}
This   is the basis of the famous {Peierls substitution}~\cite{Peierls:1933koz}, which computes the first order energy shift due to the impurity
  potential via the formal Taylor series expansion of the operator {$V(\widehat{\mbf x})$} acting on wavefunctions in perturbation theory of the lowest Landau
  level.
  
This story becomes more interesting when magnetic sources are included~\cite{Jackiw:1984rd,Gunaydin:1984isj,Mickelsson:1985fa}. 
More generally, electrically charged particles in $\bbr^3$ experience a magnetic field $\mbf B$ (possibly with
  sources) via the Lorentz force 
  \begin{align} \label{eq:Lorentz_force}
  \dot{\mbf  p}\= \frac e{m\,c} \, {\mbf
     p}\times {\mbf B} +\mbf\nabla V(\mbf x)
     \end{align}
      for the \emph{kinematical}  momentum (as opposed to the canonical momentum) $$\mbf p\=m\,\dot{\mbf x} \ , $$ where now $\mbf x=(x^1,x^2,x^3)$ and $\mbf p=(p_1,p_2,p_3)$. 
     
In the quantum theory, the Hamiltonian  $$\mathsf{H}\= \frac1{2m}\, {\mbf  p}^2+V(\mbf x)$$
  generates the Lorentz force \eqref{eq:Lorentz_force} through the Hamilton equations of motion $$-\ii\hbar \, \dot{\mbf p}\=
  [\mathsf{H},{\mbf p}]$$ only for \emph{noncommuting momenta} with the phase space commutation relations
\begin{align}\label{eq:commrelB}
[x^i,x^j]&\=0 \notag \ , \\[4pt] [x^i, p_j]&\=\ii\hbar\, \delta^i{}_j
\ , \\[4pt] [ p_i, p_j]&\= \frac{\ii \hbar\, e}c \,
\epsilon_{ijk}\, B^k \ . \notag
\end{align}
Here $\epsilon_{ijk}$ is the three-dimensional Levi-Civita symbol with $\epsilon_{123}=+1$.

The background magnetic field $\mbf B$ breaks translation invariance, but the theory has a symmetry under magnetic translations which are generated by the physical momentum $\mbf p$. A finite magnetic translation by a vector $\mbf a\in\bbr^3$ is implemented by the unitary operator
$$U({\mbf a})\=\exp\Big(\frac\ii\hbar\, {\mbf a}\cdot{\mbf
     p}\Big) \ .$$ 
The noncommutativity of momentum space is reflected in the property that the operators $\{U(\mbf a)\}_{\mbf a\in\bbr^3}$  do not form a mutually commuting set: from \eqref{eq:commrelB} we compute their commutation relations 
$$
U({\mbf a}_1)\, U({\mbf a}_2)\= \e^{\frac{\ii e}{\hbar\, c}\, \varPhi_{\mbf a_1\mbf a_2}}\, U({\mbf
  a}_1+{\mbf a}_2)\= \e^{\frac{2\ii e}{\hbar\,c}\, \varPhi_{\mbf a_1\mbf a_2}}\,U({\mbf a}_2)\, U({\mbf a}_1) \ ,
$$
where
$$\varPhi_{\mbf a_1\mbf a_2} \= \int_{\langle\mbf a_1,\mbf a_2\rangle} \, \mbf B\cdot\dd\mbf\varSigma$$ is the magnetic flux through the triangle $\langle\mbf a_1,\mbf a_2\rangle$ spanned by the vectors $\mbf a_1,\mbf a_2\in\bbr^3$.

In general, the magnetic field in fact induces \emph{nonassociating} momenta. From the commutation relations \eqref{eq:commrelB} we find that the Jacobiator of momentum components is
$$
[ p_1,  p_2, p_3 ] \ := \ [ p_1,[ p_2, p_3]] +
[ p_2,[ p_3, p_1]] + [ p_3,[ p_1, p_2]] \= \frac {\hbar^2\,e}c \,
\mbf\nabla\cdot{\mbf B}  \ . $$
Correspondingly, the magnetic translation operators do not associate,
\begin{align} \label{eq:nonassUa}
\big(U({\mbf a}_1)\, U({\mbf a}_2) \big)\, U({\mbf a}_3 )\= \e^{\frac{\ii
  e}{\hbar\,c} \, \varPhi_{\mbf a_1\mbf a_2\mbf a_3}}\, U({\mbf a}_1)\, \big(U({\mbf a}_2) \, U({\mbf
  a}_3) \big) \ ,
\end{align}
where $$\varPhi_{\mbf a_1\mbf a_2\mbf a_3}\=\int_{\langle\mbf a_1,\mbf a_2,\mbf a_3\rangle} \, \mbf\nabla\cdot \mbf B \ \dd^3\mbf x$$ is the magnetic charge enclosed by the tetrahedron $\langle\mbf a_1,\mbf a_2,\mbf a_3\rangle$ spanned by the vectors $\mbf a_1,\mbf a_2,\mbf a_3\in\bbr^3$.

In the standard Maxwell theory of electromagnetism, there are no magnetic sources and the magnetic field  $\mbf B$ is a solenoidal vector field on $\bbr^3$: $$\mbf\nabla\cdot{\mbf B}\=0 \ . $$ There are no magnetic charges anywhere, $$\varPhi_{\mbf a_1\mbf a_2\mbf a_3}\=0 \ , $$ and associativity is restored. Then $$\mbf B\=\mbf\nabla\times\mbf A$$ has a globally defined magnetic vector potential $\mbf A$ and the kinematical momentum operators in Schr\"odinger polarisation are represented by the differential operators
\begin{align}\label{eq:Schrp}
\widehat{\mbf p}\=-\ii\hbar\, \mbf\nabla-\frac ec\, {\mbf A} \ .
\end{align}
In this case, the magnetic translation operators $U(\mbf a)$ form a projective representation of the translation group of $\bbr^3$, where the projective phase is a two-cocycle determined by the magnetic flux $\varPhi_{\mbf a_1\mbf a_2}$. In particular, this recovers the simple situation of the Landau problem described at the beginning of this subsection.
  
On the other hand, when magnetic sources are present, $$\mbf\nabla\cdot{\mbf B} \ \neq\   0 \ , $$ the quantum theory exhibits 
  nonassociativity \eqref{eq:nonassUa}, unless $$\displaystyle\frac{e\,\varPhi_{\mbf a_1\mbf a_2\mbf a_3}}{2\pi\,\hbar\,c} \ \in \ \bbz$$ for all $\mbf a_1,\mbf a_2,\mbf a_3\in\bbr^3$. This is only possible for a collection of pointlike magnetic monopoles, as it would be incompatible with distributions of magnetic charge that allow for continuous deformations. In this way we recover the famous {Dirac quantization
  condition} for electric and magnetic charge, as originally observed by Jackiw~\cite{Jackiw:1984rd}.
  
\subsection{Phase Space Quantum Mechanics}
\label{sub:phasespacequantum}

In \S\S\ref{sub:magneticcharge} we arrived at a seemingly bizarre and paradoxical observation: Continuous distributions of magnetic charge require a \emph{nonassociative} formulation of quantum mechanics. In this case, the failure of the magnetic translation operators to form a projective representation is a phase factor that forms a three-cocycle of the translation group $\bbr^3$ determined by the magnetic charge $\varPhi_{\mbf a_1\mbf a_2\mbf a_3}$, which is the coboundary of the two-cochain associated to the magnetic flux $\varPhi_{\mbf a_1\mbf a_2}$. A completely analogous feature emerges in the `dual' problem for the helicity operator of massless particles~\cite{Savvidy:2025rqt}.

However, the  usual operator-state formulation of quantum mechanics, as developed by Schr\"odinger, Heisenberg, Dirac, Born and others, cannot handle nonassociative structures: Operators acting on a separable Hilbert space necessarily associate with each other. Thus we need an alternative formulation of quantum theory which does not rely on operator algebras.

An alternative approach to quantum theory was developed by Groenewold~\cite{Groenewold:1946kp} and Moyal~\cite{Moyal:1949sk} in the 1940s, following earlier work of Weyl~\cite{Weyl:1927vd}, Wigner~\cite{Wigner:1932eb} and
von~Neumann~\cite{vNeumann:1931rtl} (see~\cite{Zachos:2005gri} for a review). This approach is known as the phase space formulation of quantum mechanics. Although this
approach to quantum theory is equivalent to the more familiar approach based on canonical quantization and operator algebras, its setup is very different. Let us highlight a few of the  departures:
\begin{itemize}

\item The formalism treats position and momentum on equal footing,  and so
preserves duality symmetry between position and momentum space, unlike the usual operator realisations of phase space commutation relations such as the Schr\"odinger polarisation.

\item There are no operators or Hilbert space: Observables $O$ and states $\psi$ are described as (real) functions on phase space.

\item The algebraic structure is recovered by multiplying functions according to a star-product $$O_1\star O_2 $$ which is  noncommutative, while the trace of an operator $$\Tr\,O$$ is given by integration of $O$ over phase space.

\item A state function $S_\rho$ encodes a ``density matrix'' and is defined by 
\begin{align*}
S_\rho&\=\psi^*\star\psi \ , \\[4pt] 
\Tr\, S_\rho&\=1 \ .
\end{align*}
 
\item Expectation values of operators are given by $$\langle {\cal O}\rangle \= \Tr\,\mathcal{O}\star S_\rho \ . $$
 
\item States evolve dynamically through  the Schr\"odinger equation $$\ii\hbar\,\frac{\partial\psi}{\partial t}\= {\sf H}\star \psi \ , $$
where $\sf H$ is the classical Hamiltonian of the system.

\end{itemize}

A nonassociative star-product, based on the algebraic structure in the presence of magnetic charge, was developed in~\cite{Mylonas:2012pg} and used in~\cite{Mylonas:2013jha} to give a novel and physically sensible version of nonassociative quantum mechanics in the phase space formulation. It satisfies all the basic criteria of a quantum theory, such as reality of measurable quantities and positivity of probabilities, and reduces to standard quantum mechanics in the absence of magnetic charge. It also leads to novel effects which have no analogues in the conventional theories, such as minimal volume uncertainties which may be measurable in table-top experiments~\cite{Szabo:2017yxd}. These algebraic structures were recently abstracted into a model independent version of nonassociative quantum mechanics in~\cite{Schupp:2023jda}, which even provides an analogue of the  Gel'fand-Naimark-Segal (GNS) construction, as well as other facets of the algebraic approach to quantum theory.

\section{Noncommutative Geometry in String Theory}
\label{sec:strings}

The quantum dynamics of strings provides a natural framework in which noncommutative geometry emerges in a precise and controlled way. As a theory of quantum gravity, this will highlight the way in which string theory provides a tractable example of a quantum theory on quantized spacetime. Starting from some generalities about the spacetime geometry probed by strings and the noncommutative geometry of the closed string sector, we shall consider the open string sector and how its low energy limit naturally results in a noncommutative field theory. The model for open string dynamics that we describe is a direct extension of the simple quantum mechanics models considered in \S\ref{sec:quantumonquantized}, while our formulation of the purely stringy symmetry provided by T-duality rests on the key mathematical concepts of noncommutative geometry introduced in \S\ref{sec:formalNCG}. In the remainder of this contribution we use natural units in which $\hbar=c=1$ for simplicity. 

\subsection{String Geometry: A Prelude}

The fundamental degrees of freedom in string theory are strings, which may be closed or open. They are extended objects that have an intrinsic length scale $\ell_s$. As such, they are non-local and so see geometry in vastly different ways than the point particles of quantum field theory do. This is best exemplified through the purely stringy  symmetry of  {\it T-duality}. Let us look at the simplest example of this.

Consider  string theory on a spacetime which is compactified on a circle $\mathbbm{S}^1$ of radius $R$. As in the case of a quantum field, a string field has quantized momentum modes along the circle which are particle-like. On the other hand, the finite length $\ell_s$ of a closed string allows it to wrap around the cycle and the closed string field also has quantized winding modes along the circle, which has no analogue in quantum field theory. The quantum spectrum of the string is invariant under interchange of momentum modes with winding modes accompanied by inversion of the radius of the circle as
\begin{align}\label{eq:TdualityR}
R \ \longmapsto \ \frac{\alpha'}R \ ,
\end{align}
where $$\alpha'\=\ell_s^2$$ is the string slope.
This is the simplest example of a T-duality transformation, which is a string symmetry of spacetime that has no conventional description in classical Riemannian geometry: large and small circles are the same in string theory.

In string theory, spacetime geometry  emerges from the closed string sector and is only an approximate notion. In the setting of the circle compactification, it is valid at sizes $R\gg \ell_s$, but breaks down at scales $R\sim \ell_s$ due to non-locality of the string; note that this is near the fixed point of the T-duality transformation \eqref{eq:TdualityR} which defines the self-dual radius $R=\ell_s$. Thus we should isolate geometry from non-locality:  Geometry makes sense in the decoupling limit $\alpha'\to 0$ with $R$ finite. The zero slope limit is the limit in which massive string modes (for instance winding modes) decouple and string theory reduces to a field theory. Thus we should use effective field theories as probes of spacetime geometry. 
 
In this correlated limit, not all spacetime geometries describe ordinary geometric spaces. For example, noncommutative spaces can arise as decoupling limits, so that the effective field theories retain some of the non-locality of string theory; as in the example of a charged particle in a magnetic field, the noncommutative description is very natural and provides a conceptual as well as computationally useful framework for the quantum dynamics. The purpose of this section is to detail a concrete and controlled instance in which these statements can be made precise, through open string analogues of the effective theories of charged particles in background magnetic fields which we studied in~\S\ref{sec:quantumonquantized}. This is achieved by introducing Dirichlet branes (D-branes) in a Neveu-Schwarz $B$-field background and taking the decoupling limit, thus resulting in  noncommutative worldvolume gauge theories.

In the closed string sector, T-duality can be understood geometrically as isometries in the framework of spectral geometry discussed in \S\S\ref{sub:spectral_description}. Given a torus background with a constant Neveu-Schwarz $B$-field, one can construct a spectral triple from the closed string vertex operator algebra acting on the string Fock space, along with the Dirac-Ramond operator~\cite{Frohlich:1993es,Frohlich:1995mr,Chamseddine:1997ki,Lizzi:1997em,Lizzi:1997xe,Song:1998tr,Lizzi:1998jng,Roggenkamp:2003qp}. Then T-duality of such torus backgrounds is realised as an isomorphism between spectral triples. The isomorphism was understood in terms of Morita equivalence from \S\S\ref{sub:Morita}  in~\cite{Landi:1998ii}, where the spectral triple description was also extended to provide a manifestly noncommutative T-duality  covariant formulation of closed string theory in the low energy limit. This was cast more recently into the modern language of doubled geometry and double field theory by~\cite{Freidel:2017wst,Freidel:2017nhg,Kodzoman:2023kcm}. In the following we will be interested in how T-duality is similarly realised in the open string sector, as well as beyond the cases of constant $B$-fields.
 
\subsection{Open String Dynamics in $B$-Fields}
\label{sub:openBfields}

The non-linear sigma-model for a string propagating in target space $M$, with Riemannian metric $g$ and two-form Neveu-Schwarz field $B$, is a theory of  maps $x:\Sigma\longrightarrow M$ from an oriented Riemann surface with metric $h$  to spacetime. It is described by the action
\begin{align} \label{eq:nonlinearsigma}
S\=\int_\Sigma\,\dd^2\sigma \, \sqrt{h} \ \Big(\frac1{4\pi\,\alpha'}\,g_{ij}\,\partial x^i\cdot\partial x^j - \frac\ii2\, B_{ij}\,\partial x^i\wedge\partial x^j\Big) \ .
\end{align}
At tree-level in open string perturbation theory, the worldsheet 
$\Sigma$ is a disk or the upper complex half-plane with the standard flat metric. Then the $B$-field term is a topological term which is formally analogous to the coupling of open strings to a background magnetic field. We assume that $B$ is non-degenerate for simplicity here (and hence that the dimension $p$ of $M$ is even).

The simplest and best understood case is when $M=\bbr^{p}$ with constant metric and $B$-field~\cite{Douglas:1997fm,Ardalan:1998ce,Chu:1998qz,Schomerus:1999ug,Seiberg:1999vs,Sheikh-Jabbari:1999krr}. The two-form $B$ induces mixed boundary conditions on the fields of the non-linear sigma-model \eqref{eq:nonlinearsigma}, and in this case the two-point function on the boundary of the disk at insertion points $t,t'\in\partial\Sigma\simeq\bbr$ is given by
\begin{align} \label{eq:boundaryprop}
\big\langle x^i(t)\, x^j(t')\big\rangle \= -\alpha' \, G^{ij}
\log(t-t')^2 + \mbox{$\frac{\ii}2$} \, \theta^{ij}\, {\text{sgn}}(t-t') \ .
\end{align}
The open string metric $G$ and bivector $\theta$ are given by an explicit open-closed  string relation which determines them in terms of the closed string quantities $(g,B)$ through~\cite{Seiberg:1999vs}
\begin{align}\label{eq:openclosedgen}
\frac1{g+2\pi\,\alpha'\, B} \= \frac1G +
    \frac\theta{2\pi\, \alpha'} \ .
    \end{align}

We are  interested in this relation in the correlated decoupling limit $\epsilon\to0$ in which the string slope scales as $\alpha'\sim\epsilon^{1/2}$ and the closed string metric as $g_{ij}\sim\epsilon$, which we refer to as the Seiberg-Witten scaling limit~\cite{Seiberg:1999vs}. In this limit the relation \eqref{eq:openclosedgen} reduces to
\begin{align} \label{eq:openclosedmap}
G&\=-(2\pi\,\alpha')^2\, B\,\frac1g\,B \notag \ , \\[4pt] \theta&\=\frac1B \ .
\end{align}
In this limit, the bulk kinetic term involving the metric $g$ in the non-linear sigma-model action \eqref{eq:nonlinearsigma} vanishes and the topological $B$-field term drops to a boundary action by Stokes' Theorem. Thus the closed string modes are completely decoupled from the open string modes and the worldsheet theory described by  \eqref{eq:nonlinearsigma} becomes formally identical to the theory of charged particles in the lowest Landau level described by the Lagrangian \eqref{eq:calL0}. As in \S\S\ref{sub:magneticcharge}, we thus expect the decoupled open string dynamics to describe a quantum theory on a noncommutative space. In the following we shall make a precise statement.

These considerations extend to curved backgrounds $M$ and non-constant $B$, possibly with $H$-flux $H=\dd B\neq0$~\cite{Cornalba:2001sm,Herbst:2001ai}. Concrete examples are provided by D-branes on group manifolds, described by open string dynamics in Wess-Zumino-Witten (WZW) models~\cite{Alekseev:1999bs,Alekseev:2000fd}, and by holographic duals to certain integrable
deformations of ${\text{AdS}}_5\times \mathbbm{S}^5$ string sigma-models~\cite{vanTongeren:2015uha,vanTongeren:2016eeb,Araujo:2017jkb,Meier:2023kzt,Meier:2023lku,Borsato:2025jre}.

\subsection{The Moyal-Weyl Star-Product}
\label{sub:MWproduct}

The flurry of activity in noncommutative field theory which emanated from the late 1990s arose from the observation that, in the Seiberg-Witten scaling limit described in \S\S\ref{sub:openBfields}, the boundary propagator \eqref{eq:boundaryprop} depends only on the ordering of the points $t$ and $t'$. It is essentially independent of the worldsheet insertion points of the string fields, and hence describes a genuine spacetime quantity determined by the bivector $\theta$. For constant $B$-field, it follows that the open string interactions in tachyon scattering amplitudes are completely captured by the Moyal-Weyl star-product of functions $f,g\in C^\infty(M)$, which is given by
\begin{align} \label{eq:MoyalWeylstar}
f\star g \= \boldsymbol\cdot\, \exp\big(\mbox{$\frac{\ii}2$}\,
\theta^{ij}\, \partial_i\otimes\partial_j\big)(f\otimes g) \ ,
\end{align}
where $\partial_i=\frac\partial{\partial x^i}$. 

The formal expression \eqref{eq:MoyalWeylstar} is understood by expanding the bidifferential operator using the Maclaurin series of the exponential function, letting each leg of the tensor product act on the corresponding function in each term of the series, and then taking the pointwise product $\boldsymbol\cdot\,(f\otimes g) = f\cdot g$ of the result. Explicitly, it is given by the formal power series
\begin{align}\nonumber
f\star g &\= \sum_{n=0}^\infty \, \frac{\ii^n}{2^n\,n!} \, \theta^{i_1j_1}\cdots\theta^{i_nj_n} \, \partial_{i_1}\cdots\partial_{i_n} f\cdot\partial_{j_1}\cdots\partial_{j_n}g \\[4pt]&\= f\cdot g + \tfrac\ii2\,\theta^{ij}\,\partial_if\cdot\partial_jg +\cdots \ , \label{eq:MoyalWeylstarexpansion} 
\end{align}
showing that the star-product is a deformation of the commutative pointwise product, with leading correction proportional to the Poisson bracket $$\{f,g\}_\theta\=\theta^{ij}\,\partial_if\cdot\partial_jg$$ determined by the  bivector $\theta$. This Poisson bracket arises as the canonical bracket of zero modes induced by the boundary symplectic structure on the open string phase space, and the Moyal-Weyl product arises from the quantization of open string theory in the constant $B$-field background.

In particular, the star-commutator of coordinate functions
\begin{align}\label{eq:Heisenberg}
\big[x^i\stackrel{\star}{,}x^j\big] \ := \ x^i\star x^j - x^j\star x^i \= \ii\theta^{ij}
\end{align}
recovers the algebraic structure \eqref{eq:MoyalWeylcomm} of spacetime quantization. The Moyal-Weyl star-product is associative but noncommutative, and it provides an explicit example of the star-products discussed in \S\S\ref{sub:phasespacequantum}. When $M=\bbr^p$, the algebra $C^\infty(M)$ with the star-product \eqref{eq:MoyalWeylstar} is regarded as the algebra of functions on a noncommutative \emph{Moyal space}, denoted by $\bbr_\theta^p$.
More generally, when the $B$-field is not constant, a noncommutative star-product quantizes the bivector $\theta$ in this sense, as we shall discuss more extensively in \S\ref{sec:defquant}, in the spirit of the correspondence principle of quantum mechanics.

In physics applications one usually restricts star-products to Schwartz functions, so that in particular $f\star g$ is integrable. On this class of functions, there is also a convergent formula for the Moyal-Weyl star-product given by the convolution formula
\begin{align*}
(f\star g)(x) \= \frac1{(\pi\det\theta)^p} \, \int_{M\times M} \, \dd^py \, \dd^pz \ f(x+y)\,g(x+z) \e^{-2\ii y\cdot\theta^{-1}\,z} \ ,
\end{align*}
whose asymptotic expansion in $\theta$ coincides with \eqref{eq:MoyalWeylstarexpansion}.

\subsection{Noncommutative Yang-Mills Theory}
\label{sub:NCYM}

Open strings attach to D-branes, which are hyperplanes in spacetime, that here we will take to be the entire $p$-dimensional target space $M$ of the non-linear sigma-model \eqref{eq:nonlinearsigma}, with a slight abuse of notation. D-branes act as sources of spacetime curvature and interact gravitationally with one another through exchanges of closed strings. The massless bosonic modes in the open string sector are gauge fields $A_i$ on the worldvolume $M$ of the D-branes (coming from Neumann boundary conditions for the non-linear sigma-model \eqref{eq:nonlinearsigma}), and scalar fields $X^a$ on $M$ describing the transverse fluctuations of the D-branes in spacetime (coming from Dirichlet boundary conditions); geometrically, these are described by a gauge connection $A$ together with an adjoint section $X$ of a Hermitian vector bundle on $M$, called the Chan-Paton bundle of the D-branes. This connection between geometry and gauge theory forms the basis of string dualities such as the AdS/CFT correspondence. Since the Seiberg-Witten scaling limit decouples the closed string sector, the 
 low-energy dynamics is described
  by a field theory on the D-branes which is decoupled from gravity. For non-zero $B$-field, this is a deformation of  gauge theory called \emph{noncommutative gauge theory}. 
  
Restricting to the pure gauge sector, the low energy effective action for a stack of $N$ parallel coincident D-branes wrapping $M$ is the action of noncommutative $U(N)$ Yang-Mills theory
\begin{align}\label{eq:NCYM}
I_{\text{\tiny YM}} \= \frac1{2g_{\text{\tiny YM}}^2} \, \int_M\,\dd^px \, \sqrt{\det G} \ G^{ik}\,G^{jl}\,\Tr\,F_{ij}\star F_{kl} \ ,
\end{align}
where  $$
F_{ij}\=\partial_i A_j-\partial_j
A_i-\ii[A_i\stackrel{\star}{,}A_j] $$
is the noncommutative field strength tensor and here  $\star$ means the tensor product of the star-product \eqref{eq:MoyalWeylstar}  with matrix multiplication. The effective Yang-Mills coupling constant $g_{\text{\tiny YM}}$ in the D$p$-brane gauge theory is related to string slope $\alpha'$ and the closed string coupling $g_s$ through the combination~\cite{Seiberg:1999vs}
\begin{align}\label{eq:YMcoupling}
g_{\text{\tiny YM}}^2 \= \frac{(2\pi)^{p-2}}{(\alpha')^{(3-p)/2}} \ g_s \, \e^\phi \
\bigg(\frac{\det(g+2\pi\,\alpha'\, B)}{\det g}\bigg)^{1/2} \ ,
\end{align}
where $\phi$ is the dilaton. This remains finite in the decoupling limit if $g_s\e^\phi\sim\epsilon^{(3-p+r)/4}$, where $r$ is the rank of $B$ (most of the time we assume full rank $r=p$).

The action \eqref{eq:NCYM} is invariant under infinitesimal noncommutative gauge (or star-gauge) transformations
\begin{align}\label{eq:stargauge}
\delta_\lambda^\star A_i \= \partial_i\lambda+\ii[\lambda\stackrel{\star}{,}A_i] \ ,
\end{align}
for Hermitian gauge parameters $\lambda\in C^\infty(M,\mathfrak{u}(N))$.
Noncommutative Yang-Mills theory actually first appeared  much earlier in the mathematics literature:  Connes and Rieffel considered the classical theory in the late 1980s using an operator algebraic approach when $M=\bbT^2$ is a two-dimensional torus~\cite{Connes:1987ue,Rieffel:1990eb}.

From the perspective of the worldsheet quantum field theory, the derivation of the low energy effective action from open string scattering amplitudes is strongly dependent on the choice of regularisation scheme: In Pauli-Villars regularization one arrives at an ordinary gauge theory with complicated non-local higher derivative interactions in which the dependence on the $B$-field is explicit, whereas in point-splitting regularization one arrives at a noncommutative gauge theory in which the non-locality and $B$-field dependence is absorbed into the star-product $\star$, i.e. into the noncommutativity of spacetime. Since the target space physics should be independent of the chosen worldsheet regularisation scheme, this suggests that noncommutative Yang-Mills theory should be ``dual'' to some ordinary gauge theory.

That such a duality exists and can be worked out explicitly through a field redefinition is arguably the most remarkable result of the seminal paper~\cite{Seiberg:1999vs}. The field redefinition $A\longmapsto\widehat A[A]$ from ordinary gauge theory to noncommutative gauge theory is called the \emph{Seiberg-Witten map}, where we use hats to distinguish noncommutative fields and their star-gauge transformations from their ordinary counterparts. The Seiberg-Witten map only makes sense as a map between gauge orbits of fields
\begin{align}\nonumber
\widehat A[A+\delta_\lambda A]\=\widehat A[A] + \widehat{\delta}_{\widehat\lambda[\lambda,A]}\widehat A[A] \ .
\end{align}
This can be solved as an asymptotic series in $\theta$ whose leading terms read
$$
\widehat{A}{\,}^i[A,\theta]\=\theta^{ij}\,A_j+\tfrac12\,\theta^{kl}\,A_l\,\big[\partial_k(\theta^{ij}\,A_j)-\theta^{ij}\,F_{jk}\big] + \cdots \ ,
$$
with a similar asymptotic expansion for the noncommutative gauge parameters $\widehat\lambda[\lambda,A]$, which are  field dependent.

The Seiberg-Witten map implies that noncommutative Yang-Mills theory can be cast as an ordinary gauge theory, which however is \emph{not} the usual Yang-Mills theory; in particular, it retains the non-locality of the noncommutative theory which ``remembers'' that it orginates from string theory.
This dual commutative description of noncommutative gauge theory led to a plethora of investigations in all sorts of directions, together with associated phenomenological implications of noncommutativity (which we do not attempt to review or even summarise here). The duality makes sense in fact for arbitrary (non-constant) Poisson bivector $\theta$. We shall give a geometric description of the Seiberg-Witten map in~\S\ref{sec:defquant}.

\subsection{Morita Duality of Noncommutative Gauge Theory}
\label{sub:openstringTduality}

Let us now return to T-duality and descibe its fate in the Seiberg-Witten scaling limit, as well as its realisation in the low energy effective theory. We compactify the target space coordinates so that $M=\bbr^p$ is the universal cover of a $p$-dimensional torus $\bbT^p$, on which  $x^i$ are now local coordinates. D-branes carry charges under higher-form Ramond-Ramond fields in string theory. Since D-branes wrapping $\bbT^p$ come with a Chan-Paton gauge bundle $E\longrightarrow\bbT^p$, these topological charges  are classified by the K-theory group $K(\bbT^p)$ of vector bundles on $\bbT^p$. In the presence of a constant $B$-field and in the Seiberg-Witten scaling limit, the algebra of functions $C^\infty(\bbT^p)$ is deformed to a noncommutative algebra by the Moyal-Weyl star-product \eqref{eq:MoyalWeylstar}, which we think of as the algebra of functions on a \emph{noncommutative torus}, denoted by $\bbT^p_\theta$. Then the Chan-Paton bundle becomes a projective module over $\bbt_\theta^p$ and the K-theory group $K(\bbt^p_\theta)$ classifies D-brane charges in the $B$-field background.

In the absence of topologically non-trivial $H$-flux, the T-dual of the torus $\bbt^p$ is the dual torus $\widehat\bbt{}^p$ by the Buscher rules. T-duality along a single one-cycle interchanges Neumann and Dirichlet boundary conditions in the non-linear sigma-model \eqref{eq:nonlinearsigma}, so interchanges D$p$-branes and D($p\pm1)$-branes. Open string T-duality on a $p$-torus $\bbT^p$ acts on D$p$-brane charges in the following way. The action of the T-duality group $SO(p,p;\bbz)$ on the closed string variables $(g,B)$ can be expressed through the combination $$\bun\=\frac1{\alpha'}\, \big(g+2\pi\,\alpha'\,B\big)$$  as a generalized linear fractional transformation sending $\bun\longmapsto\bun'$ where $$\displaystyle{\bun'\=(A\,\bun+B)\, \frac1{C\,\bun+D}} \quad , \quad \begin{pmatrix} A & B \\ C &
    D \end{pmatrix} \ \in \ SO(p,p;\bbz) \ . $$
    
Using the open-closed string relation in the decoupling limit \eqref{eq:openclosedmap}, this acts on the open string variables as $(G,\theta)\longmapsto(G',\theta')$ where
\begin{align*}
G'&\=(C\,\theta+D)\,G\,(C\,\theta+D)^\top \ , \\[4pt]
\theta'&\=(A\,\theta+B)\,\frac1{C\,\theta+D} \ ,
\end{align*}
and on the Yang-Mills coupling \eqref{eq:YMcoupling} as $g_{\text{\tiny YM}}\longmapsto g_{\text{\tiny YM}}'$ where $$g_{\text{\tiny YM}}' \= g^{\phantom{\dag}}_{\text{\tiny YM}} \, \big|\det(C\,\theta+D)\,\big|^{1/4} 
 \ . $$ 
This $SO(p,p;\bbz)$-action is defined on the dense subset of the space of antisymmetric matrices $\theta$ where $C\,\theta+D$ is invertible.

It is a remarkable feature of noncommutative gauge theory that it inherits this T-duality symmetry~\cite{Seiberg:1999vs}: the action \eqref{eq:NCYM} is invariant under these combined transformations. This is unlike the situation in ordinary quantum field theory where the action of T-duality would have no meaning: the non-locality kept by the Seiberg-Witten scaling limit yields a field theory that is invariant under the stringy T-duality symmetry. This provides a geometric refinement of \emph{topological T-duality} which coincides with Morita equivalence of the noncommutative tori $\bbt_\theta^p$ and \smash{$\bbt_{\theta'}^p$}. In particular, there is an identification of corresponding K-theory groups $$
K\big(\bbT_\theta^p\big) \= K\big(\bbT_{\theta'}^p\big)
\ , $$ 
which is the underlying topological manifestation of the T-duality symmetry of noncommutative Yang-Mills theory. See e.g.~\cite{Connes:1997cr,Schwarz:1998qj,Rieffel:1998vs,Brace:1998ku,Brace:1998xz,Pioline:1999xg} for further details and developments.

\subsection{Morita Duality and Non-Geometric Backgrounds}
\label{sub:nongeometric}

The story thus far becomes more interesting and new features appear for backgrounds with topologically non-trivial $H$-flux, that is, when $$H\=\dd B \ \neq \ 0$$ represents a non-trivial cohomology class. In these cases ``non-geometric'' backgrounds can arise after T-duality~\cite{Hull:2004in,Shelton:2005cf,Dabholkar:2005ve}, which have no rigorous formulation in conventional geometry. 

The prototypical situation starts with a spacetime $M$ which is a principal torus bundle of rank $p$
\begin{align}\nonumber
\xymatrix{\bbT^p \ \lhook\!\!\joinrel\ar[r] & \ M \ar[d] \\ & \ \mathbbm{S}_x^1}
\end{align}
over a circle $\mathbbm{S}_x^1$ with local coordinate $x\in[0,1)$.
For a non-trivial fibration, the fibres $\bbt^p$ carry a non-trivial monodromy group action: they patch together by large diffeomorphisms in $GL(p,\bbz)$ under a periodic shift $x\longmapsto x+1$.

Suppose now that $M$ carries a non-zero $H$-flux. In string theory, fluxes are quantized due to generalised Dirac charge quantization so that they have integer periods. Whence the three-form $H$ is a de~Rham representative of an integer cohomology class $$[H] \ \in \ H^3(M,\bbz)$$  which corresponds to the characteristic class of a gerbe on $M$, called its Dixmier-Douady invariant. In certain situations, applying a fibrewise T-duality to $M$ results in a space which is still fibred over the original circle $\mathbbm{S}_x^1$, but is no longer a torus bundle: the fibres $\bbt^p$  now patch together by T-duality transformations in $O(p,p;\bbz)$  under  $x\longmapsto x+1$. This is called a \emph{T-fold}: it is a perfectly well-defined background in string theory, but globally it has no formulation as a classical geometric space; in particular the dual closed string variables $(g',B')$ are only locally defined due to their dependence on the base coordinate~$x$.
 
Can the T-fold be described rigorously? Noncommutative geometry provides precisely such a description through the formalism of topological T-duality~\cite{Mathai:2004qq}  in the open string sector. The original torus bundle with $H$-flux is the spectrum
$$
M\= {\text{Spec}}(\CA)
$$
of a continuous trace algebra $\CA$ whose Dixmier-Douady invariant is represented by the three-form $H$; the corresponding K-theory group $K(\CA)$ classifies D-brane charges in the $H$-flux background. As a $C(M)$-module algebra, $\CA$ can be identified with the sections of a locally trivial algebra bundle $\mathscr{E}\longrightarrow M$ with fibre $\CK$ and structure group $$\text{Aut}(\CK) \= \text{Inn}(\CK)\=\text{Ad}\big(\CU(\CK)\big) \ , $$ where $$\CK\=\CK(\hil)$$ is the algebra of compact operators (i.e. the limits of finite rank operators) on a separable infinite-dimensional Hilbert space~$\hil$. 

Fibrewise T-duality corresponds to an action $$\alpha:\bbr^p \ \longrightarrow \ {\text{Aut}}(\CA)$$
of the abelian Lie group $\bbr^p$ by automorphisms of $\CA$, which is induced from the $\bbt^p$-action on the torus bundle. The T-dual background is then described as the spectrum of the crossed product algebra
$$\widehat\CA\= \CA\rtimes_\alpha
\widehat\bbr{}^p \ , $$ where $\widehat\bbr{}^p:=\text{Hom}_{\mathscr{Ab}}(\bbr^p,U(1))\simeq\bbr^p$ is the Pontryagin dual of $\bbr^p$ in the category of abelian groups $\mathscr{Ab}$. The algebra $\widehat\CA$ is Morita equivalent to $\CA$, and in particular there is an identification of K-theory groups $$K(\widehat\CA\,)\=K(\CA) \ . $$

In the situations described above, this provides a rigorous description of a T-fold as an algebra bundle of noncommutative tori
\begin{align}\nonumber
\xymatrix{\bbT_{\theta(x)}^p \ \lhook\!\!\joinrel\ar[r] & \ \widehat\CA \ar[d] \\ & \ \mathbbm{S}_x^1}
\end{align}
wherein the non-geometric fibrewise monodromies become Morita equivalences of noncommutative tori. That is, $\theta(x+1)$ is an $SO(p,p;\bbz)$-orbit of $\theta(x)$ for each $x\in[0,1)$, and the noncommutative fibre torus $\bbT^p_{\theta(x+1)}$ is Morita equivalent to $\bbT^p_{\theta(x)}$; in particular, noncommutative Yang-Mills theory is well-defined on the T-fold \smash{$\widehat\CA$}. See e.g.~\cite{Mathai:2004qq,Brodzki:2006fi,Brodzki:2007hg,Bouwknegt:2008kd,Aschieri:2020uqp} for further details and developments.

As an explicit illustration of this formalism, let us look at the simplest example~\cite{Ellwood:2006my,Grange:2006es,Hull:2019iuy}. Consider a three-dimensional torus $\bbt^3$ with constant $H$-flux, viewed as a trivial  torus bundle $$\bbt^3\=\mathbbm{S}_x^1\times\bbt^2$$ of rank two over a circle. The metric on $\bbt^3$ is the flat metric
\begin{align}\nonumber
\dd s^2\=(2\pi\,r\, \dd x)^2+\frac \ttA{\tau_2^\circ} \, \big|\,\dd y^1+ \tau^\circ\, \dd y^2\,\big|^2 \ ,
\end{align}
where $r>0$ is the radius of $\mathbbm{S}_x^1$, while $\ttA>0$ is the area and $\tau^\circ=\tau_1^\circ+\ii\tau_2^\circ\in\bbc$ with $\tau_2^\circ>0$  the complex structure modulus of the fibre $\bbt^2$ with local coordinates $(y^1,y^2)\in[0,1)^{\times 2}$. The Neveu-Schwarz fields are
\begin{align}\notag
B&\=n\, x\, \dd y^1\wedge \dd y^2 \ , \\[4pt]
H&\=\dd B\=n \, \dd x\wedge \dd y^1\wedge\dd y^2 \ , \notag
\end{align}
where $n=[H]\in H^3(\bbt^3,\bbz)=\bbz$, while the dilaton field vanishes. 

The T-duality group of string theory compactified on the fibre $\bbt^2$ is
\begin{align}\notag
O(2,2;\bbz) \ \simeq \ \big(SL(2,\bbz)_{\tau}
\times SL(2,\bbz)_\rho \big) \rtimes \big(\bbz_2 \times\bbz_2\big) \ .
\end{align}
The first $SL(2,\bbz)$ factor is the mapping class group of $\bbt^2$ which acts geometrically through M\"obius transformations of $\tau^\circ$, while the second acts on the complexified K\"ahler modulus
\begin{align}\notag
\rho \= B + \frac\ii{2\pi\,\alpha'} \, \ttA \ .
\end{align}
This has a non-trivial monodromy
\begin{align}\notag
\CM \= \Big({\small\begin{matrix} 1 & n \\ 0 & 1 \end{matrix} } \normalsize \Big)
\end{align}
of infinite order in a parabolic conjugacy class of $SL(2,\bbz)_\rho$ which generates
\begin{align}\notag
\rho(x+1) \= \CM\big[\rho(x)\big]\=\rho(x)+n \ .
\end{align}
The monodromy shifts the $B$-field by the two-form $n\,\dd y^1\wedge\dd y^2$, which represents an integral cohomology class in $H^2(\bbt^3,\bbz)=\bbz\oplus\bbz\oplus\bbz$.

Let us now wrap a D1-brane around the base one-cycle $\mathbbm{S}_x^1$ in this three-torus. Via the Buscher rules, T-duality along the abelian isometries generated by the global vector fields $\frac\partial{\partial y^1}$ and $\frac\partial{\partial y^2}$ maps the D1-brane to a D3-brane filling a T-fold whose closed string variables $(g,B,\phi)$ depend on the modulus
\begin{align}\notag
\tau(x) \= \tau^\circ+n\, x \ .
\end{align}
This has a non-trivial monodromy
\begin{align}\notag
\tau(x+1)\=\CM\big[\tau(x)\big]\=\tau(x)+n
\end{align}
in a parabolic conjugacy class of $SL(2,\bbz)_\tau$. The K\"ahler modulus of the $\bbt^2$ fibres
\begin{align}\notag
\rho\=\frac1{\overline{\tau}(x)}
\end{align}
exhibits the background as a T-fold with monodromy
\begin{align}\notag
\rho(x+1)\=\frac{\rho(x)}{1+n\,\rho(x)}
\end{align} 
in $SL(2,\bbz)_\rho$.

Applying the open-closed string transformation \eqref{eq:openclosedgen} along the fibre directions gives the open string variables $(G,\theta)$ seen by the D3-brane. This yields a worldvolume of topology $\mathbbm{S}_x^1\times\bbt^2$.
It is shown in~\cite{Hull:2019iuy} that there is a consistent scaling limit with $\alpha',\ttA,\tau_2^\circ,g_s\sim\epsilon^{1/2}$ such that as $\epsilon\to 0$ the open string parameters on the D3-brane become finite quantities
\begin{align}\notag
G&\= \big(2\pi\,r_1 \, \dd y^1\big)^2
+ \big(2\pi\,r_2 \, \dd y^2\big)^2 \ , \\[4pt]
\theta &\= n\,x \ \frac\partial{\partial y^1} \wedge\frac\partial{\partial y^2} \notag \ , \\[4pt] 
g^2_{\text{\tiny YM}}&\=2\pi\,\bar g_s \ . \notag
\end{align}
The open string metric $G$ is just the usual flat metric on a rectangular two-torus $\bbt^2$ of constant radii $r_1$ and $r_2$, while the open string bivector $$\theta(x)\=n\,x$$ depends on the base coordinate of the worldvolume $\mathbbm{S}_x^1\times\bbt^2$.

Since $\frac\partial{\partial y^i}\theta=0$ for $i=1,2$, the Kontsevich quantization formula of \S\S\ref{sub:Kontsevich} below gives the
star-product of functions on the D3-brane worldvolume \smash{$\mathbbm{S}_x^1\times \bbt^2$} as the Moyal-Weyl type formula
$$
f\star g \= \, \mbf\cdot\, \exp\left[\, \frac\ii2 \, n\, x \,
\left(\frac\partial{\partial y^1} \otimes \frac\partial{\partial y^2} - \frac\partial{\partial y^2}
\otimes \frac\partial{\partial y^1}\right) \right] (f\otimes g) \ .
$$
This  quantizes the three-dimensional Heisenberg algebra
\begin{align}\notag
\big[y^1\stackrel{\star}{,}y^2\big] &\= \ii n\,x \ , \\[4pt] \big[y^1\stackrel{\star}{,}x\big] \=\big[y^2\stackrel{\star}{,}x\big] &\= 0 \ . \notag
\end{align} 
Via Fourier transformation it agrees on $\mathbbm{S}_x^1\times \bbt^2$ with the product in the twisted Schwartz convolution algebra $C^*(G_\bbz)$, where $G_\bbz$ is a lattice inside the three-dimensional Heisenberg group $G_\bbr$. The group $C^*$-algebra  $C^*(G_\bbz)$ is stably isomorphic to the bundle of noncommutative two-tori $\bbt_{\theta(x)}^2$ over $\mathbbm{S}_x^1$ obtained from topological T-duality~\cite{Mathai:2004qq,Brodzki:2006fi,Aschieri:2020uqp}
$$\widehat\CA\= C^*(G_\bbz)\otimes\CK \ . $$

The noncommutative torus $\bbt^2_{\theta(x)}$ has Morita equivalence group $$SO(2,2;\IZ) \ \simeq \ \big(SL(2,\IZ)_{\theta}\times SL(2,\IZ)_\tau\big)\,\big/\,\bbz_2 \ . $$
The monodromies give the correct parabolic $SL(2,\bbz)_\theta$ Morita transformations
$$
\theta(x+1) \= \CM\big[\theta(x)\big] \= \theta(x)+n \ .
$$
Thus in this example open strings see a conventional geometric three-torus $\bbt^3$, instead of a T-fold, with non-geometric
  noncommutativity $\theta(x)$. The noncommutative gauge theory on the D3-brane is perfectly well-defined: the Morita equivalence symmetry
  of noncommutative Yang-Mills theory is inherited from T-duality in the decoupling limit.

\section{Deformation Quantization}
\label{sec:defquant}

We will now consider some formal aspects of the formulation of noncommutative field theories, with the aim of going beyond the cases considered in \S\ref{sec:strings}. We look at two particular schemes for deformation quantization, each of which comes with its own merits and applications. The first scheme enables a more precise geometric definition of the Seiberg-Witten map discussed in \S\S\ref{sub:NCYM}. The second scheme allows for a more precise description of the notion of quantum symmetry discussed in \S\ref{sec:spacetimequant}. More detailed treatments of these aspects of deformation quantization can be found in the contributions of Pierre Bieliavsky and Marija Dimitrijevi\'c \'Ciri\'c to this volume. 

\subsection{Kontsevich Formality}
\label{sub:Kontsevich}

Kontsevich famously proved that any Poisson manifold $(M,\theta)$ admits a deformation quantization~\cite{Kontsevich:1997vb}. The basis of Kontsevich's construction is the Formality Theorem: there is a sequence of formality maps $U_n$ for $n\in\mathbbm{N}_0$ from multivector fields to multidifferential operators, which define an $L_\infty$-quasi-isomorphism of differential graded Lie algebras relating Schouten brackets $[-,-]_S$ to
  Gerstenhaber brackets $[-,-]_G$.
  
In particular, when $M=\bbr^p$, the star-product $\star$ which quantizes a bivector $\theta$ is given by a formal power series
\begin{align} \label{eq:Kontstar}
f\star g \= \widehat\varPsi(\theta)(f,g) \ := \ \sum_{n=0}^{\infty}\, {\frac{\ii^n}{n!}\,
   U_{n}(\theta, \dots ,\theta )} (f,g)
   \end{align}
for functions $f,g\in C^\infty(M)$. 
Each term in this series has a graphical representation which can be computed using a precise combinatorial prescription as integrals on the upper complex half-plane. 

This may be represented as the Feynman diagram expansion of a topological string theory with target space $M$~\cite{Cattaneo:1999fm}, the Poisson sigma-model, using path integral techniques and the Batalin-Vilkovisky (BV) formalism, which connects the Kontsevich formula to the open string dynamics discussed in \S\S\ref{sub:openBfields}. The action of the Poisson sigma-model is~\cite{Ikeda:1993fh,Schaller:1994es}
\begin{align}\label{eq:PsigmaM}
S_{\text{\tiny P}\sigma\text{\tiny M}} \= \int_{\Sigma}\, \Big(\xi_i\wedge \dd x^i+\frac12\,
\theta^{ij}(x)\, \xi_i\wedge \xi_j \Big) \ .
\end{align}
When $$\theta\=\frac1B \ , $$ integrating out the worldsheet one-forms $\xi_i$ shows that this action is equivalent on-shell to the string $B$-field coupling
\begin{align*}
\int_{\Sigma} \, x^*(B) \= \frac12\, \int_{\Sigma}\, B_{ij}(x)\, \dd x^i\wedge\dd x^j
\end{align*}
in the non-linear sigma-model \eqref{eq:nonlinearsigma}.

The formality maps are required to satisfy a set of formality conditions. The simplest of these reads as
$$[\widehat\varPsi(\theta),\star]_G\= \ii \widehat\varPsi([\theta,\theta]_S) \ ,$$
where here $\star$ is regarded as a bidifferential operator.
This implies that the star-product $\star$ is associative, $$[\widehat\varPsi(\theta),\star]_G\=0 \ , $$ if and only if $\theta$ is a Poisson bivector, $$[\theta,\theta]_S\=0 \ . $$ This is the original setting of Kontsevich's construction and it arises as the on-shell condition (BV master equation) for the Poisson sigma-model \eqref{eq:PsigmaM}. 

However, as noted by~\cite{Cornalba:2001sm,Mylonas:2012pg}, the formula \eqref{eq:Kontstar} still makes sense more generally beyond the case of Poisson structures and can serve to define a nonassociative star-product. In this case the non-vanishing Schouten bracket $$[\theta,\theta]_S \ \neq \ 0$$ implies that the Jacobi identity for the bracket $$\{f,g\}_\theta\=\theta^{ij}\,\partial_if\cdot\partial_jg$$ is violated. This is related to the presence of $H$-flux $$H\=\dd B \ \neq \ 0$$ in the string theory setting, and is formally analogous to the nonassociativity in quantum mechanics with magnetic monopoles that we discussed in \S\ref{sec:quantumonquantized}.
 
Up to gauge equivalence, the leading terms in the expansion \eqref{eq:Kontstar} are given by
\begin{align}\label{eq:Kontstarleading}
\begin{split}
f\star g  \= f\cdot g & + \tfrac{\ii}2\,{\theta^{ij}\,\partial_if\cdot\partial_jg} -\tfrac{1}4\,\theta^{ij}\,\theta^{kl}\,\partial_i\partial_kf\cdot\partial_j\partial_lg + \cdots \\
& -\tfrac{1}6\,{\theta^{ij}\,\partial_j\theta^{kl}}\,(\partial_i\partial_kf\cdot\partial_lg - \partial_kf\cdot\partial_i\partial_lg) + \cdots \ .
\end{split}
\end{align}
In the first line of \eqref{eq:Kontstarleading} we recognise the expansion of the Moyal-Weyl star-product \eqref{eq:MoyalWeylstarexpansion}, while the second line involves specific combinations of derivatives of the bivector $\theta$. In particular, in the cases when 
\begin{align}\label{eq:MoyalWeylcondition}
{\theta^{ij}\,\partial_j\theta^{kl}\=0}
\end{align} 
the series \eqref{eq:Kontstarleading} exponentiates just like the Moyal-Weyl formula \eqref{eq:MoyalWeylstar}. Note that the condition \eqref{eq:MoyalWeylcondition} holds for a very broad class of  bivectors which are not necessarily constant, such as the Poisson bivector we encountered in \S\S\ref{sub:nongeometric}.

\subsection{Seiberg-Witten Maps}
\label{sub:SWmaps}

We will now give an exact all-orders description of the Seiberg-Witten map discussed in~\S\S\ref{sub:NCYM}~\cite{Jurco:2000fs}. Here we work with an arbitrary Poisson bivector $\theta$ and $U(1)$ gauge theory for simplicity. Let $A$ be a $U(1)$ gauge field and $$F\=\dd A$$ its curvature. 

Let us start by recalling the classical Moser Lemma from symplectic geometry. This states that the map from a non-degenerate closed two-form $B$ to a ``nearby'' $B$-field $$B'\=B+F$$ is a coordinate change: it is given by the flow $\varphi$ generated by the vector field $$\xi\=\theta^{ij}\,A_j\,\frac\partial{\partial x^i} \ , $$
where
\begin{align}\notag
\theta\=\frac1B \ .
\end{align}
This maps the bivector $\theta$ to
\begin{align}\label{eq:thetaMoser}
\theta'\=\theta\, \frac1{1+F\, \theta} \ .
\end{align}
The transformation \eqref{eq:thetaMoser} makes sense for arbitrary (possibly degenerate) Poisson structures.

By applying the Kontsevich formality maps to this classical construction, we arrive at a quantum Moser Lemma where the flow $\varphi$ is quantized to a covariantizing map $\cald$: the image $\widehat\varPsi(\xi)$ defines a deformed diffeomorphism which integrates to a  ``flow'' $\cald$ given by an invertible differential operator. Let $\star$ (resp. $\star'$) be the Kontsevich star-product which quantizes the bivector $\theta$ (resp.~$\theta'$). Then the covariantizing map is an equivalence of these star-products, in the sense that
$$\cald(f\star g)\=\cald f\star'\cald
  g \ ,$$
  for all $f,g\in C^\infty(M)$.
We may summarise the quantum Moser Lemma by the commutative diagram 
$$
\xymatrix{\theta \ \ar[d]_\varphi \ar[r]^{\widehat\varPsi}  & \ \star
  \ar[d]^\cald \\ \theta' \ \ar[r]_{\widehat\varPsi}  & \ \star'}
$$

Noncommutative $U(1)$ gauge fields $\widehat A$ are now introduced through the \emph{covariant coordinates}
$$X_i\=\theta^{-1}_{ij}\,\cald x^i \ =: \ \theta^{-1}_{ij}\,x^j+\widehat A_i \ . $$ 
The significance of this definition is that \eqref{eq:Kontstarleading} implies that derivatives can be implemented by the adjoint actions of coordinate functions through
\begin{align}\nonumber
[x^i\stackrel{\star}{,}f]\=\ii\theta^{ij}\,\partial_jf \ ,
\end{align}
for any function $f\in C^\infty(M)$. Quantization then sends an ordinary infinitesimal $U(1)$ gauge transformation
$$\delta_\lambda A\=\dd\lambda$$
to the covariant transformation $$\widehat\delta_{\widehat\lambda}\cald f\=\ii[\,\widehat\lambda\stackrel{\star}{,}\cald f] \ . $$

In particular, this implies the noncommutative $U(1)$ gauge transformation $$\widehat{\delta}_{\widehat\lambda}\widehat A\=\dd\widehat\lambda+\ii[\,\widehat\lambda\stackrel{\star}{,}\widehat A\,] \ . $$
{Thus in the covariant coordinate formalism, noncommutative gauge transformations are realised as inner automorphisms of the star-product algebra.} This is precisely the perspective on gauge symmetries that we introduced in \S\S\ref{sub:Morita}.

\subsection{Hopf Algebras and Twist Quantization}
\label{sec:Drinfeldtwist}

Let us now switch perspectives and discuss a more algebraic approach to deformation quantization based on cocycle twists of Hopf algebras; see the classic text~\cite{Majid:1996kd} for an elaborate account. Let $\mathbbmss{H}$ be a  quasi-triangular Hopf algebra over $\bbc$ with structure maps $({\mbf\Delta},\mathsf{S},\mathsf{R},\varepsilon,\cdot)$, where $\,\cdot\,$ is the multiplication on $\mathbbmss{H}$, and the coproduct $\mbf{\Delta}:\mathbbmss{H}\longrightarrow \mathbbmss{H}\otimes \mathbbmss{H}$, the antipode $\mathsf{S}:\mathbbmss{H}\longrightarrow \mathbbmss{H}$, and the counit $\varepsilon:\mathbbmss{H}\longrightarrow \bbc$ are unital algebra (anti)homomorphisms, while $\mathsf{R}\in \mathbbmss{H}\otimes \mathbbmss{H}$ is the universal R-matrix.

A \emph{Drinfel'd twist} for $\mathbbmss{H}$ is an invertible element
$$
\mathsf{F}\=\mathsf{F}_{(1)}\otimes \mathsf{F}_{(2)} \ \in \ \mathbbmss{H}\otimes \mathbbmss{H} \ ,
$$
where we use a notation similar to the standard Sweedler convention (with implicit summation understood). It is required to satisfy two conditions that make it a counital two-cocycle on $\mathbbmss{H}$: the \emph{cocycle condition}
\begin{align}\label{eq:cocyclecond}
(\mathsf{F}\otimes 1)\,(\mbf{\Delta}\otimes\text{id})(\mathsf{F})\= (1\otimes \mathsf{F})\, (\text{id}\otimes\mbf{\Delta})(\mathsf{F})
\end{align}
in $\mathbbmss{H}\otimes \mathbbmss{H}\otimes \mathbbmss{H}$, and the \emph{counitality condition}
\begin{align}\nonumber
(\varepsilon\otimes\text{id})(\mathsf{F})\=1\=(\text{id}\otimes\varepsilon)(\mathsf{F} )
\end{align}
in $\mathbbmss{H}$.

A Drinfel'd twist $\mathsf F$ maps $\mathbbmss{H}$ to a new quasi-triangular Hopf algebra
  $\mathbbmss{H}_\mathsf{F}$ with the same underlying vector space as $\mathbbmss{H}$ and structure maps $(\mbf{\Delta}_\mathsf{F},\mathsf{S}_\mathsf{F},\mathsf{R}_\mathsf{F},\varepsilon,\cdot)$. The formulas for the twisted coproduct, antipode and universal R-matrix are
\begin{align*}
\mbf{\Delta}_\mathsf{F}(\xi)&\= \mathsf{F}\, \mbf{\Delta}( \xi)\, \mathsf{F}^{-1} \ , \\[4pt]
 \mathsf{S}_\mathsf{F}(\xi)&\=\chi\,\mathsf{S}(\xi)\,\chi^{-1} \ , \\[4pt]
  \mathsf{R}_\mathsf{F}&\=\mathsf{F}^{-1}_{21}\,\mathsf{R}\,\mathsf{F} \ ,
\end{align*}
where $\xi\in \mathbbmss{H}$, while $$\chi\=\mathsf{F}_{(1)}\,\mathsf{S}(\mathsf{F}_{(2)})$$ and $$\mathsf{F}_{21}=\mathsf{F}_{(2)}\otimes \mathsf{F}_{(1)} \ . $$
In Sweedler notation we write
\begin{align}\nonumber
\mbf{\Delta}_\mathsf{F}(\xi)\=\xi_{(1)}\otimes\xi_{(2)} \ .
\end{align}

The twisting procedure quantizes any left $\mathbbmss{H}$-module algebra  $\CA$  to an algebra  $\CA_\mathsf{F}$ with the same underlying vector space as $\CA$, but with the product $\,\cdot\,$ on $\CA$ deformed to the associative 
  noncommutative product
$$
f \star g\= \boldsymbol\cdot\, \big(\mathsf{F}^{-1}\triangleright(f\otimes g)\big)\= \big(\mathsf{F}^{-1}_{(1)}\triangleright f\big) \cdot \big(\mathsf{F}^{-1}_{(2)}\triangleright g\big) \ ,
$$
for $f,g\in\CA$. Here we write $\triangleright:\mathbbmss{H}\otimes\CA\longrightarrow\CA$ for the action of $\mathbbmss{H}$ on $\CA$. Then $\CA_\mathsf{F}$ carries a representation of the twisted Hopf algebra $\mathbbmss{H}_\mathsf{F}$, making it into a left $\mathbbmss{H}_\mathsf{F}$-module algebra:
\begin{align}\nonumber
\xi\,\triangleright(f\star g) & \= \boldsymbol\cdot\, \big(\mbf{\Delta}(\xi)\,\mathsf{F}^{-1}\triangleright(f\otimes g)\big) \\[4pt] \label{eq:modulealgebra}
& \= \boldsymbol\cdot\, \big(\mathsf{F}^{-1}\,\mbf{\Delta}_\mathsf{F}(\xi)\triangleright(f\otimes g)\big) \\[4pt]
& \= (\xi_{(1)}\triangleright f)\star(\xi_{(2)}\triangleright g) \ , \notag
\end{align}
where we use the same symbol $\triangleright$ for the action of $\mathbbmss{H}_\mathsf{F}$ on $\CA_\mathsf{F}$.

Note that associativity of $\star$ is guaranteed by the cocycle condition \eqref{eq:cocyclecond}, and if $\alg$ is a commutative algebra then $\alg_\mathsf{F}$ is a \emph{braided}-commutative (or \emph{quasi}-commutative) algebra:
\begin{align}\nonumber
f\star g \= (\mathsf{R}_{\mathsf{F}(1)}^{-1}\triangleright g)\star(\mathsf{R}_{\mathsf{F}(2)}^{-1}\triangleright f) \ ,
\end{align}
where
\begin{align*}
\mathsf{R}_\mathsf{F} \= \mathsf{R}_{\mathsf{F}(1)}\otimes \mathsf{R}_{\mathsf{F}(2)} \ .
\end{align*}
 
In fact, the twisting procedure simultaneously quantizes all  $\mathbbmss{H}$-covariant constructions: there are braided monoidal categories ${}^\mathbbmss{H}\mathscr{M}$ and ${}^{\mathbbmss{H}_\mathsf{F}}\mathscr{M}$ of left $\mathbbmss{H}$-modules and left $\mathbbmss{H}_\mathsf{F}$-modules, respectively, and twisting gives a braided monoidal 
  functorial isomorphism
$$
\mathscr{Q}_\mathsf{F}\,:\, {}^\mathbbmss{H}\mathscr{M}~ \longrightarrow~ {}^{\mathbbmss{H}_\mathsf{F}}\mathscr{M} \ .
$$

More generally, by dropping the cocycle condition \eqref{eq:cocyclecond}, we replace a Drinfel'd twist by a cochain twist $\mathsf{F}\in \mathbbmss{H}\otimes \mathbbmss{H}$ which maps $\mathbbmss{H}$ to a quasi-Hopf algebra $\mathbbmss{H}_\mathsf{F}$. The coboundary of $\mathsf{F}$, $$\mbf{\Phi}_\mathsf{F}\=\mbf\partial^*\mathsf{F}\=\mbf{\Phi}_{\mathsf{F}(1)}\otimes\mbf{\Phi}_{\mathsf{F}(2)}\otimes\mbf{\Phi}_{\mathsf{F}(3)} \ \in \ \mathbbmss{H}_\mathsf{F}\otimes \mathbbmss{H}_\mathsf{F}\otimes \mathbbmss{H}_\mathsf{F} \ , $$ is a counital three-cocycle on the quasi-Hopf algebra $\mathbbmss{H}_\mathsf{F}$. A cochain twist $\mathsf F$ quantizes any $\mathbbmss{H}$-module algebra $\CA$  to a ``quasi-associative'' $\mathbbmss{H}_\mathsf{F}$-module algebra $\CA_\mathsf{F}$, in the sense that associativity of the deformed product $\star$ holds only up to the action of the three-cocycle~$\mbf{\Phi}_\mathsf{F}$:
\begin{align}\nonumber
(f\star g)\star h \= (\mbf{\Phi}_{\mathsf{F}(1)}\triangleright f)\star\big((\mbf{\Phi}_{\mathsf{F}(2)}\triangleright g)\star(\mbf{\Phi}_{\mathsf{F}(3)}\triangleright h)\big) \ .
\end{align}  
Nonassociative geometry is developed in this setting in~\cite{Beggs:2005zt,Barnes:2014ksa,Barnes:2015uxa,Blumenhagen:2016vpb,Aschieri:2017sug}. It is precisely such a three-cocycle, associated to magnetic translations, that is responsible for the nonassociativity in quantum mechanics with magnetic monopoles that we discussed in \S\ref{sec:quantumonquantized}.

\subsection{Twisted Symmetries and Noncommutative Gravity}
\label{sub:twistedsymmetries}

A common situation in both geometry and field theory starts with the universal enveloping Hopf algebra $$\mathbbmss{H} \= \mathbbmss{U}\frg$$  of a Lie algebra $\frg$ of infinitesimal symmetries of a manifold $M$. The structure maps are defined on primitive elements $X\in\frg$ by
\begin{align*}
\mbf\Delta(X)&\=X\otimes 1 + 1 \otimes X \ , \\[4pt]
 \mathsf{S}(X)&\=-X \ , \\[4pt]
  \varepsilon(X)&\=0 \ ,
\end{align*}
and extended as unital (anti)homomorphisms. The triangular structure is given by the trivial R-matrix
\begin{align}\notag
\mathsf{R}\=1\otimes 1 \ .
\end{align}
Then as $\mathbbmss{U}\frg$-module algebras $\CA$ one can take the algebra of smooth functions $C^\infty(M)$, the differential graded exterior algebra of differential forms $\Omega^\bullet(M)$, and more generally tensor fields on $M$. 

A particularly important example for the structure of quantum spacetimes is the enveloping algebra $$\mathbbmss{H} \=  \mathbbmss{U}\Gamma(TM)$$ of vector fields on $M$, generating diffeomorphisms of $M$. If $\alg$ is a $\mathbbmss{U}\Gamma(TM)$-module algebra of the type mentioned above, then  the Lie algebra $\Gamma(TM)$ acts on $\alg$ via the Lie derivative and the Leibniz rule. From this it is possible to construct a twisted differential calculus on $M$ with the ordinary (undeformed) de~Rham differential $\dd$.

For example, when $M=\bbr^p$ and $\theta$ is a constant bivector, the canonical example of a Drinfel'd twist is provided by the Moyal-Weyl twist
\begin{align}\label{eq:MWtwist}
\mathsf{F}\=\exp\big(-\tfrac\ii2\,\theta^{ij}\,\partial_i\otimes\partial_j\big) \ .
\end{align}
With this twist, deformation quantization of the algebra of functions $\CA=C^\infty(M)$ reproduces the Moyal-Weyl star-product \eqref{eq:MoyalWeylstar}. In this example the braiding is given by the triangular R-matrix 
$$\mathsf{R}_\mathsf{F}\=\mathsf{F}^{-2} \ . $$ 

A great virtue of the twist approach to deformation quantization  is that it provides simultaneous characterisations of both a quantum space and its quantum symmetries, with the latter encoded by the twisted Hopf algebra $\mathbbmss{H}_\mathsf{F}$. A \emph{twisted symmetry} acts on $\mathbbmss{H}_\mathsf{F}$-modules through the twisted coproduct $\mbf{\Delta}_\mathsf{F}$ (see e.g.~\cite{Szabo:2006wx,Aschieri:2009zz} for reviews). They coincide with the usual symmetries when acting on single-particle states, but differ in their action on multi-particle states. For example, consider an infinitesimal outer automorphism $\delta_\xi:\CA\longrightarrow\CA$, parametrized by $\xi\in \mathbbmss{H}$, of an $\mathbbmss{H}$-module algebra $\CA$. From the calculation \eqref{eq:modulealgebra} it follows that its twisted action on $\CA_\mathsf{F}$ is given by
$$
\delta_\xi(f\star g) \= \delta_{\xi_{(1)}}f\,\star\,\delta_{\xi_{(2)}}g \ .
$$
This is different from the way a classical symmetry would act, which in general also transforms the twist $\mathsf{F}$ and hence the star-product: here $\xi\in \mathbbmss{H}$  does not ``see'' the star-product. Thus symmetries which are otherwise broken because they do not preserve a bivector $\theta$, e.g.~Lorentz transformations in the case of the Moyal-Weyl twist \eqref{eq:MWtwist}, can be realised instead as twisted symmetries.

Let us now try to apply this principle to the string theory setting discussed in \S\ref{sec:strings}, focusing on the closed string sector in the spacetime $M=\bbr^p$. Contrary to the open string sector, the massless bosonic modes of closed strings give background geometry and gravity determined by a  metric $g$, a Kalb-Ramond two-form field $B$, and a dilaton $\phi$. These modes are treated as couplings in the non-linear sigma-model, and worldsheet conformal invariance implies that the beta-functions for these couplings vanish. These in turn can be interpreted as target space equations of motion for $g$, $B$ and $\phi$ which at one-loop order derive from the Neveu-Schwarz part of the closed string effective action (in  Einstein frame)
$$
I_{\rm closed}\=\int_M\,\dd^px\,\sqrt{\det g} \ \Big( \text{Scal}_g-\frac1{12}\,\e^{-\phi/3} \, |H|_g^2 - \frac16\,(\nabla\phi)^2\Big) \ ,
$$
where $\text{Scal}_g$ is the scalar curvature of $g$ and $H=\dd B$. 

Is there a  noncommutative version of this gravity theory, akin to the noncommutative Yang-Mills theory of \S\S\ref{sub:NCYM}, and in particular a duality with ordinary gravity similar to that provided by the Seiberg-Witten map? To answer these questions, one may appeal to the formulation of general relativity on noncommutative spacetime which was developed in the seminal works~\cite{Aschieri:2005yw,Aschieri:2005zs}. In this approach, a noncommutative tensor calculus with twisted diffeomorphism symmetry leads to a gravity theory with deformed Einstein field equations, which admit noncommutative black hole solutions~\cite{Ohl:2008tw,Ohl:2009pv,Schupp:2009pt} and exhibit many other novel features. Twisted noncommutative gravity has become a vast and active area since its inception in the mid-2000s; it is outside of the scope of this survey to review or even cite the huge body of literature on the subject. 

Going back to our motivating scenario, we only mention that twisted diffeomorphisms {\it do not} appear to arise as physical symmetries of closed string theory: the low-energy scaling limit of the induced gravitational action on a D-brane in a constant  $B$-field contains additional contributions to the three-graviton interaction vertex that cannot be derived from an action based on twisted diffeomorphism symmetry~\cite{Alvarez-Gaume:2006qrw}. While this would seem to cast a dark shadow of doubt on the physical relevance of twisted symmetries in general, we note that twisted symmetries  {\it do} arise in some settings of the AdS/CFT correspondence: certain integrable Yang-Baxter deformations of string theory on $\text{AdS}_5\times \mathbbm{S}^5$ have Drinfel’d twisted symmetry, and are conjecturally dual to noncommutative Yang-Mills theory with twisted Poincar\'e symmetries; see e.g.~\cite{Vicedo:2015pna,vanTongeren:2015uha,Borsato:2021fuy,Meier:2023kzt,Meier:2023lku,Borsato:2025jre} for further details of these developments.

\section{Noncommutative Renormalization}
\label{sec:formalQFT}

In this section we will discuss the extent to which the perturbative expansions of noncommutative field theories define sensible quantum field theories, focusing on the simplest examples of scalar fields on $\bbr^p$ interactiing through the Moyal-Weyl star-product \eqref{eq:MoyalWeylstar}. Due to non-locality of these field theories induced by the star-product, they are generally non-renormalizable when treated with standard quantization techniques. We describe two ways to restore renormalizability: We may either modify the free field theory using background magnetic fields along the lines discussed in \S\S\ref{sub:magneticcharge},  thereby explicitly breaking translation invariance of the theory, or we may modify the path integral in a way compatible with the twisted symmetries discussed in \S\S\ref{sub:twistedsymmetries}, thus restoring the full Poincar\'e symmetry. In this sense the spacetime symmetries of noncommutative field theory are intimately related to their renormalization~\cite{Szabo:2007ra}.

\subsection{UV/IR Mixing}
\label{sub:UVIRmixing}

Let us first discuss the basic underlying problem compared to conventional field theories. 
Noncommutative quantum field theories are typically plagued by the problem of ultraviolet/infrared (UV/IR) mixing~\cite{Minwalla:1999px}. To explain this, let us start from the Moyal-Weyl star-product \eqref{eq:MoyalWeylstar} of two plane waves on $\bbr^p$, which leads to the Baker-Campbell-Hausdorff formula
\begin{align}\label{eq:BCH}
\e^{\ii k\cdot x}\star\e^{\ii q\cdot x} \= \e^{\frac\ii2\, k\cdot\theta\, q} \ \e^{\ii(k+q)\cdot x} \ ,
\end{align}
where we use the short-hand notation
\begin{align}\nonumber
k\cdot\theta\, q\=
k_i\,\theta^{ij}\,q_j \ .
\end{align}
The noncommutative phase factors in \eqref{eq:BCH} appear in open string amplitudes and account for the mixing induced by a $B$-field background.

Let $\phi$ be a real scalar field on $\bbr^p$. Its Fourier transform $\widetilde\phi$ obeys the reality property $$\overline{\widetilde\phi(k)}\=\widetilde\phi(-k) \ . $$ 
The formula \eqref{eq:BCH} implies that the  star-product of $\phi$ with itself $$\phi^{\star 2}\=\phi\star\phi$$ modifies its usual Fourier convolution product $\widetilde\phi(k)\,\widetilde\phi(q)$ in momentum space to
$$
\widetilde\phi(k)\,\widetilde\phi(q)~
\e^{\frac\ii2\, k\cdot\theta\, q} \ .
$$

In particular, an interaction term $$\frac\lambda{n!}\,\phi^{\star n}$$ in the classical action yields an interaction vertex in momentum space in the form of a phase factor
\begin{align}\label{eq:phinvertex}
\frac\lambda{n!} \, (2\pi)^p~
\e^{\frac\ii2\sum\limits_{I<J}\,k_I\cdot\theta\, k_J} \ ,
\end{align}
together with momentum conservation $k_1+k_2+\cdots+k_n=0$ as a consquence of translation invariance of the noncommutative scalar field theory. The phase factor becomes effective at energies {$E$}
with {$E\,\sqrt{\theta}\ll1$}, where $$\theta \ := \ \max_{1\leqslant i,j\leqslant p}\,|\theta^{ij}| \ . $$ In this infrared region, noncommutative field theory is drastically different from conventional quantum field theory.

The vertex \eqref{eq:phinvertex} is not invariant under arbitrary permutations of the momenta $k_I$, but only under cyclic permutations due to momentum conservation. As a consequence, in conventional perturbation theory the Feynman diagrams are represented as ribbon graphs in order to keep track of the order in which lines emanate from vertices, and can be naturally organised into two classes: planar and non-planar graphs, referring to whether or not they can be drawn on a plane without crossing the fattened lines. Planar diagrams arise from contracting neighbouring legs and involve momentum space integrals which are identical to their commutative counterparts up to overall phase factors of the form \eqref{eq:phinvertex} depending only on external momenta; in particular they exhibit the same ultraviolet divergences. Non-planar graphs, on the other hand, arise from contractions of non-neighbouring legs and contain phase factors \eqref{eq:phinvertex} involving internal loop momenta. 

If we introduce an ultraviolet cutoff $\Lambda$ to regulate Feynman integrals, when  $\theta\neq0$ the non-planar diagrams generically remain  finite as $\Lambda\to\infty$ due to the rapid phase oscillations from the noncommutative interactions. But this is no longer true when external momenta $p$ become exceptional, i.e. when $\theta\,p=0$, where their amplitude grows beyond any bound: exceptional momenta render the phase factors ineffective and the original ultraviolet divergence reappears as an infrared divergence. In other words, in a non-planar diagram, an ultraviolet cutoff $\Lambda$ induces an effective infrared cutoff 
\begin{align}\nonumber
\Lambda_0\=\frac1{\theta\,\Lambda} \ .
\end{align}
In the limit $\theta\to0$ with $\Lambda$ finite, $\Lambda_0\to\infty$ and the non-planar diagrams become divergent for any external momenta. This non-analytic dependence of amplitudes on $\theta$ is called \emph{UV/IR mixing}. That low energy processes can receive contributions from high energy virtual particles appears to ruin standard renormalization schemes, such as the Wilsonian renormalization group approach, which require a clear separation of energy scales.

On its own, a one-loop non-planar diagram is generically well-defined at any non-exceptional external momenta. But when inserted as subgraphs into higher-loop graphs, these exceptional momenta are realised in loop integrations and result in severe divergences for any number of external legs. In such graphs the products of Fourier transforms of the usual distributions of quantum field theory are well-defined, while those involving the oscillatory phase factors are not.

Consider, for example, the Euclidean Feynman propagator for a real scalar field $\phi$ of mass $m$, which is the distribution given by
\begin{align}\nonumber
\varDelta(x-y)\ := \ \int_{\bbr^p}\, \frac{\dd^pq}{(2\pi)^p} \ \frac{\e^{\ii q\cdot (x-y)}}{q^2+m^2}  \ = \ \int_{\bbr^p}\, \frac{\dd^pq}{(2\pi)^p} \ \widetilde{\varDelta}(q) \ \e^{\ii q\cdot (x-y)} \ .
\end{align}
In momentum space, products of the Fourier transform $$\widetilde{\varDelta}(q)\=\frac1{q^2+m^2}$$ are always well-defined. 

However, this is not true for the distribution
\begin{align}\nonumber
\varUpsilon(x-y) \ :\!\!&= \  \int_{\bbr^p\times\bbr^p}\, \frac{\dd^pk}{(2\pi)^p} \, \frac{\dd^pq}{(2\pi)^p} \ \frac{\e^{\ii k\cdot\theta\, q}}{k^2+m^2} \, \frac{\e^{\ii q\cdot (x-y)}}{q^2+m^2} \\[4pt] \nonumber
\ &= \ \int_{\bbr^p}\, \frac{\dd^pq}{(2\pi)^p} \ \varDelta(\theta\,q) \, \widetilde{\varDelta}(q) \ \e^{\ii q\cdot (x-y)} \ ,
\end{align}
which arises, for example, as the one-loop non-planar contribution to the two-point function in $\phi^{\star 4}$-theory.
In momentum space, the Fourier transform
\begin{align}\nonumber
\widetilde{\varUpsilon}(q) \= \varDelta(\theta\,q) \, \widetilde{\varDelta}(q)
\end{align}
contains both $\widetilde\varDelta$ and the distribution $\varDelta$ itself.

In $p$ dimensions, products  $\varDelta(x)^n$ of the distribution $\varDelta$ with itself are ill-defined for $n\geqslant p-2$, due to the usual ultraviolet singularity at $x=0$. In products of $\widetilde \varUpsilon$ with itself this singularity appears as an infrared singularity at $\theta\,q=0$. If $\varUpsilon$ appears $p-2$ or more times in a graph, it leads to an uncontrollable divergence. These divergences increase with the order of  perturbation theory, and all correlation functions are affected and diverge. {We conclude that the field theory \emph{cannot} be renormalized}.

The mixing of degrees of freedom implied by this problem have various physical interpretations and consequences. Foremost is the fact that interactions in noncommutative field theory are \emph{non-local}: If $\phi$ and $\psi$ are fields supported in a region of
  small size $\ell\ll\sqrt{\theta}$, then their star-product 
  {$\phi\star\psi$} is non-zero in a large region of size
  {$\theta/\ell$}. The Baker-Campbell-Hausdorff formula \eqref{eq:BCH} together with Fourier transformation of arbitrary fields $\phi(x)$ on $\bbr^p$ give
$$\e^{\ii k\cdot x}\star\phi(x)\star\e^{-\ii k\cdot
    x}\=\phi(x-\theta\,k) \ . $$
This exhibits the non-locality of the theory: the infrared dynamics are governed by quanta which behave like non-local ``dipoles'' with electric dipole moment $$\Delta
  x^i\=\theta^{ij}\,k_j \ . $$ They are analogous to electron-hole bound states in a strong magnetic field, like in the Landau problem discussed in \S\S\ref{sub:magneticcharge}. The dipoles interact by joining at their ends~\cite{Sheikh-Jabbari:1999cvv,Bigatti:1999iz}.

On the other hand, the ultraviolet dynamics are governed by the elementary quantum fields $\phi$ themselves, which create quanta with
  pointlike momenta {$k_i$}, since in this region the effects of noncommutativity are negligible. It was suggested in~\cite{Rey:2002vy} that there is a certain UV/IR ``duality'' which relates dynamics in the two regimes: UV/IR mixing is then due to the asymmetry between supports of fields on extended and pointlike degrees of freedom in the different regimes. We will discuss an explicit realization of this duality in \S\S\ref{sub:GWmodel} below.

UV/IR mixing has phenomenological applications, see e.g.~\cite{Amelino-Camelia:2001hzq,Helling:2007zv}. In the infrared regime, it breaks translation invariance leading to modified dispersion relations of the form
$$
E^2\=\mbf p^2+m^2+\Delta M^2\Big(\frac1{\theta\,p}\Big) \ .
$$
This can be compared with experiments for energies $E$ in the range $$
\Lambda_0 \ < \ E \ < \ \Lambda\=\frac1{\theta\,\Lambda_0} \ .
$$   

Noncommutative field theories with sufficient supersymmetry do not suffer from the problem of UV/IR mixing due to cancellations between bosonic and fermionic loops. In the following we discuss two modifications of standard noncommutative field theory which cure the UV/IR mixing problem without supersymmetry.

\subsection{UV/IR Duality}
\label{sub:GWmodel}

We start by seeking a covariant version of the quantum field theory which makes ultraviolet and infrared regimes indistinguishable, and hence makes the UV/IR ``duality'' discussed in \S\S\ref{sub:UVIRmixing} into a true symmetry. The idea of duality covariant noncommutative field theory was originally implemented in a simple physical model of charged particles in constant  magnetic fields, interacting non-locally through the star-product~\cite{Langmann:2002cc,Langmann:2003cg,Langmann:2003if}. Covariantization turns the ultraviolet degrees of freedom into extended objects by replacing their pointlike momenta with ``Landau'' momenta
$$
k_i~\longmapsto~K_i\=k_i-\tfrac12\,B_{ij}\,x^j \ ,
$$
where $B_{ij}$ is a constant ``magnetic'' background, generically independent of $\theta^{ij}$. In canonical quantization, this generates a 
``noncommutative momentum space'' $$[K_i,K_j]\=-\ii
  B_{ij} \ . $$
There is now a symmetry $$k_i \ \longleftrightarrow \ -\tfrac12\,B_{ij}\,x^j$$ between position and momentum spaces, and hence no distinction between what is meant by ultraviolet or infrared.

To implement this idea analytically, let $\psi(x)$ be a complex scalar field of mass $m$ on Euclidean space $\bbr^p$ with quartic noncommutative interaction. The action is
\begin{align}\label{eq:Scov}
S_{\rm cov} \= \int_{\bbr^p} \, \dd^px \ \Big[\psi^\dag\star\big(\mathsf{H}_B+m^2\big)\,\psi + \frac g2\,\psi^\dag\star\psi\star\psi^\dag\star\psi\Big] \ ,
\end{align}
where
\begin{align}\label{eq:LandauHam}
\mathsf{H}_B \= -\nabla_i\,\nabla^i
\end{align}
is the ``Landau'' Hamiltonian with the covariant derivative
\begin{align}\nonumber
\nabla_i\=\partial_i-\tfrac\ii2\,B_{ij}\,x^j
\end{align}
for a constant non-degenerate background $B_{ij}=-B_{ji}$. The curvature of the connection $\nabla$ is given by
\begin{align}\nonumber
[\nabla_i,\nabla_j]\=-\ii B_{ij} \ ,
\end{align}
thus exhibiting a noncommutative momentum space exactly as in the quantum mechanical model of \S\S\ref{sub:magneticcharge}.

It is instructive to rewrite this field theory in an operator formalism, by essentially reversing the mapping discussed in \S\S\ref{sub:phasespacequantum}. For this, we represent the spacetime commutation relations
\begin{align}\nonumber
[\widehat x{}^i,\widehat x{}^j]\=\ii\theta^{ij}
\end{align}
by Hermitian operators $\widehat x^i$ acting on a separable Hilbert space $\hil$. The algebra $\CA(\bbr_\theta^p)$ of functions on the Moyal space is represented on $\hil$ through the linear Weyl transform
\begin{align}\nonumber
f(x) \ \longmapsto \ \widehat f\=f(\widehat x\,) \ := \ \int_{\bbr^p}\,\frac{\dd^pk}{(2\pi)^p} \ \widetilde f(k) \ \exp(\ii k_i\,\widehat x{}^i) \ .
\end{align}
This is indeed a representation of $\CA(\bbr_\theta^p)$,
\begin{align}\label{eq:Weylstar}
\widehat f \ \widehat g \= \widehat{f\star g} \ ,
\end{align}
showing that the star-product is the pullback to the space of functions on $\bbr^p$ of the composition of operators on $\hil$, and in this sense quantizes the classical Poisson algebra determined by the bivector $\theta$. 
Moreover, spacetime averages of fields are represented by traces of operators on $\hil$:
\begin{align}\nonumber
\int_{\bbr^p}\,\dd^px \ f(x) \= \Tr\,(\widehat f\ ) \ .
\end{align}

We will construct this representation in a way which mimicks the definition of the Moyal-Weyl star-product \eqref{eq:MoyalWeylstar}, and consequently makes manifest the duality between noncommutative position and momentum spaces. For this, we introduce Hermitian coordinate operators $\widehat r{}^i$ on $\hil$ for $i=1,\dots,p$ which form a commuting set:
\begin{align}\label{eq:rrels}
[\widehat r{}^i,\widehat r{}^j]\=0  \ .
\end{align}
We also introduce anti-Hermitian operators $\widehat\partial_i$ on $\hil$ for $i=1,\dots,p$ which act through outer derivations on the algebra generated by $\widehat r{}^i$ as
\begin{align}
[\widehat\partial_i,\widehat\partial_j]&\=0 \ , \notag \\[4pt] [\widehat\partial_i,\widehat r{}^j]&\=\delta_i{}^j \ . \label{eq:partialrels}
\end{align}

The mutually commuting coordinate operators $\widehat r{}^i$ have a simultaneous complete system of orthonormal eigenstates $|x\rangle\in\hil$, labelled by $x\in\bbr^p$, which form the Schr\"odinger representation of the commutation relations \eqref{eq:rrels} and \eqref{eq:partialrels}: 
\begin{align*}
\widehat r{}^i|x\rangle&\=x^i\,|x\rangle \ , \\[4pt]
 |x\rangle&\=\exp(-x^i\,\widehat\partial_i)|0\rangle \ .
 \end{align*}
Ordinary fields $f(x)$ on $\bbr^p$ are then represented through superpositions of these coordinate eigenstates as
\begin{align}\nonumber
|f\rangle&\=\int_{\bbr^p}\,\dd^px \ f(x)\,|x\rangle
\ , \\[4pt]
f(x)&\=\langle x|f\rangle \ . \notag
\end{align}
In other words, the spectrum of the commutative $\bbc$-algebra generated by $\widehat r{}^i$ for $i=1,\dots,p$ is the ordinary Euclidean space $\bbr^p$.

Now we represent the noncommuting coordinates $x^i$ and covariant derivatives $\nabla_i$ on the Hilbert space $\hil$ by the Hermitian operators
\begin{align}
\widehat x{}^i&\=\widehat r{}^i+\tfrac\ii2\,\theta^{ij}\,\widehat\partial_j \notag \ , \\[4pt]
 \widehat p_i&\=-\ii\widehat\partial_i-\tfrac12\,B_{ij}\,\widehat r{}^j \ , \label{eq:hatxprep}
\end{align}
which are analogous to the familiar Bopp shift of canonical quantum mechanics.
In the Schr\"odinger representation $|x\rangle$, they coincide respectively with the operators \eqref{eq:Schrx} and \eqref{eq:Schrp} that appeared in the quantum mechanics of a charged particle in a constant magnetic field. From \eqref{eq:rrels} and \eqref{eq:partialrels} it follows that the operators \eqref{eq:hatxprep} obey the commutation relations
\begin{align} 
[\widehat x{}^i,\widehat x{}^j]&\=\ii\theta^{ij} \notag , \\[4pt] [\widehat p_i,\widehat p_j]&\=-\ii B_{ij} \label{eq:xprels}\ , \\[4pt]
 [\widehat p_i,\widehat x{}^j]&\=\ii \big(\unit-\tfrac14\,B\,\theta\big)_i{}^j \ . \notag
\end{align}
Note that the operators $\widehat p_i$ and $\widehat x{}^j$ are \emph{not} canonically conjugate unless one of the background fields $B$ or $\theta$ vanishes.

With this representation the field theory defined by \eqref{eq:Scov} becomes a ``matrix model'' with action
\begin{align}\label{eq:Scovmatrix}
S_{\rm cov}\=\Tr\,\Big[\psi(\widehat x\,)^\dag\,\big(\widehat{\mbf p}{}^2+m^2\big)\,\psi(\widehat x\,) + \big(\psi(\widehat x\,)^\dag\,\psi(\widehat x\,)\big)^2\Big] \ ,
\end{align}
where 
\begin{align}\nonumber
\widehat{\mathsf{H}}_B\,\widehat\psi\=\widehat{\mbf p}{}^2\,\widehat\psi \ := \ [\widehat p_i,[\widehat p{}^i,\widehat\psi\,]]
\end{align}
is the representation of the Landau Hamiltonian \eqref{eq:LandauHam} on the Hilbert space $\hil$. 

At the special point in the moduli space of this theory where 
\begin{align}\label{eq:selfdualpoint}
\frac\theta2\=\Big(\frac B2\Big)^{-1} \ ,
\end{align}
it follows from the commutation relations \eqref{eq:xprels} that the action \eqref{eq:Scovmatrix} has a huge symmetry under rotations of the operators $\psi(\widehat x\,)$ by the gauge group $\CU(\CA(\bbr_\theta^p))\subset U(\hil)\simeq U(\infty)$ in the unitary group of the separable Hilbert space $\hil$:
\begin{align}\nonumber
\psi(\widehat x\,) \ \longmapsto \ U(\widehat x\,)\,\psi(\widehat x\,) \ , \quad U(\widehat x\,)\ \in \ \CU\big(\CA(\bbr_\theta^p)\big) \ .
\end{align}
In the star-product formulation, it follows from \eqref{eq:Weylstar} that these unitary rotations translate into finite  ``star-gauge'' transformations
\begin{align}\nonumber
\psi(x) \ \longmapsto \ (U\star\psi)(x) \ .
\end{align}

This $U(\infty)$ symmetry corresponds to an invariance of the action \eqref{eq:Scov} under deformed canonical transformations of $\bbr^p$ with the Poisson structure $\theta=4B^{-1}$~\cite{Lizzi:2001nd}. At this point the quantum field theory is exactly solvable and resembles a ``discrete $\bbz_2$''  noncommutative gauge theory, in the spirit of the Connes-Lott model from \S\S\ref{sub:StandardModel}. We will encounter this $U(\infty)$ symmetry again in \S\S\ref{sub:twistedreduced} during our treatment of noncommutative Yang-Mills theory.

For generic moduli $(\theta,B)$, it was shown in~\cite{Langmann:2002cc} that the quantum field theory has a duality under Fourier transformation of the scalar fields. In particular, the duality sends the noncommutativity parameters $(\theta,B)$ to $(\theta^{-1},B^{-1})$ while rescaling the spacetime coordinates $x$ to $$\widetilde x\=2\,\theta^{-1}\,x \ . $$ Both the classical action \eqref{eq:Scov} and all quantum correlation functions are symmetric under this duality. The special point \eqref{eq:selfdualpoint} is the self-dual point, where the duality covariance becomes an exact invariance of the theory, and resembles the formal ``zero rank'' limit of open string T-duality of noncommutative Yang-Mills theory on a $p$-torus $\bbt^p$ that we discussed in \S\S\ref{sub:openstringTduality}.

\subsection{The Grosse-Wulkenhaar Model}

In real noncommutative $\phi^4$-theory, the analog of coupling to a magnetic background is the addition of a background harmonic oscillator potential to the standard kinetic operator $\Delta=\partial_i\,\partial^i$ as
\begin{align}\label{eq:Deltashift}
\Delta \ \longmapsto \ \Delta + \tfrac12\,\omega^2\,\widetilde x^2 \ .
\end{align}
This noncommutative field theory is known as the {\it Grosse-Wulkenhaar model}~\cite{Grosse:2004yu,Grosse:2004by,Grosse:2004wte}: The real Euclidean scalar quantum field theory with action
\begin{align}\nonumber
S_{\text{\tiny GW}} \= \int_{\bbr^p}\,\dd^px \ \Big(\frac12\,\partial_i\phi\star\partial^i\phi + \frac{\omega^2}2\,(\widetilde x_i\,\phi)\star(\widetilde x{}^i\,\phi) + \frac{m^2}2\,\phi\star\phi + \frac\lambda{4!}\,\phi\star\phi\star\phi\star\phi\Big)
\end{align}
is symmetric under the position/momentum space duality $$k_i \ \longleftrightarrow \ \widetilde x_i\=2\,\theta^{-1}_{ij}\,x^j$$ induced by Fourier transformation of fields. This covariant model is renormalizable to all orders in $\lambda$: The confining harmonic oscillator potential serves as an infrared cutoff, and is the unique one with the UV/IR duality symmetry that makes the field theory just renormalisable.
 
The remarkable renormalisability features of the Grosse-Wulkenhaar model have been studied from various points of view, see e.g.\cite{Rivasseau:2005bh,Gurau:2005gd,Disertori:2006uy,Disertori:2006nq}. It presents a novel renormalization group flow that corresponds to a new mixture of standard ultraviolet and infrared notions. For example, at the self-dual point $$\omega\=1 $$ where the quantum field theory is \emph{invariant} under the UV/IR duality, the beta-functions for the coupling $\lambda$ and frequency $\omega$ vanish:
\begin{align}\nonumber
\beta_\lambda\=\beta_\omega\=0 \ .
\end{align}
Among other things, this implies that there is no Landau ghost (renormalons), and hence a non-trivial nonperturbative completion of the Grosse-Wulkenhaar model is believed to be possible. This is in marked contrast to the situation in the usual commutative $\phi^4$ field theory.

A particularly appealing feature of the Grosse-Wulkenhaar model is that the quantum field theory can be reformulated as the $N\to\infty$ limit of an $N\times N$ matrix model in an external field, by expanding the scalar fields in the basis of Landau wavefunctions, i.e.~the eigenfunctions of the Landau Hamiltonian \eqref{eq:LandauHam} at the self-dual point \eqref{eq:selfdualpoint}. There is no analogue of such a natural matrix regularization in ordinary quantum field theory. We will return to the relation between noncommutative field theories and matrix models in \S\ref{sec:matrixmodels}.

In spite of its technical appeal, some conceptual issues surround this model, most notably the physical origin and meaning of the ``magnetic background'' $\omega$. In~\cite{Buric:2009ss} a geometric origin of the modification \eqref{eq:Deltashift} was derived in the matrix model formulation, relating the harmonic potential term to the coupling of the scalar fields with the background curvature of a noncommutative space obtained via finite-dimensional matrix truncation of the Heisenberg algebra \eqref{eq:Heisenberg}.

\subsection{Braided Quantum Field Theory}

Another  approach to renormalizable noncommutative field theory is by modifying the \emph{path integral} directly, rather than the classical theory. This is called {braided quantum field theory}. The renormalization properties of braided quantum field theory are very
  different. In particular, UV/IR mixing seems to be far less severe and maybe even
  absent~\cite{Oeckl:2000eg,Balachandran:2005pn,Bu:2006ha,Fiore:2007vg,Fabiano:2025pub}. Among other features, this approach restores full Poincar\'e invariance of noncommutative field theory through the notion of twisted symmetry discussed in \S\S\ref{sub:twistedsymmetries}, which in this framework is more natural to delineate as a `braided symmetry'.
  
A concrete rigorous way of defining these models is via Oeckl's algebraic approach to {braided quantum field theory}, which is based on braided versions of Wick's Theorem and Gaussian integration~\cite{Oeckl:1999zu,Sasai:2007me}.
  A more recent systematic approach employs the modern purely algebraic formalism of Batalin-Vilkovisky
  (BV) quantization, as developed by {Costello and Gwilliam}~\cite{Costello:2021jvx}, which avoids explicit path integral techniques altogether. The braided
  generalization of the BV formalism is dual to the  {braided ${L_\infty}$-algebras} that construct {braided field theories} which are equivariant under the action of a triangular Hopf algebra, with braided commutative fields~\cite{DimitrijevicCiric:2021jea,Giotopoulos:2021ieg,Nguyen:2021rsa,DimitrijevicCiric:2023hua,Bogdanovic:2024jnf,DimitrijevicCiric:2024qew}; in particular, it has no known path integral formulation. In all examples studied thus far, the general feature appears to be that braided quantum field theories are free from UV/IR mixing, at least in the absence of non-abelian gauge symmetries. 
  
One main advantage of the braided BV formalism is that it is naturally equipped to handle braided gauge symmetries~\cite{DimitrijevicCiric:2021jea,Giotopoulos:2021ieg,Nguyen:2021rsa}, in contrast to the approaches of Oeckl and others. Braided gauge symmetries are based on braided Lie algebras, which are Lie algebras in the braided monoidal representation category of a triangular Hopf algebra. A natural class is provided by twisting the bracket $[-,-]_\frg$ of a Lie algebra $\frg$ of infinitesimal symmetries of a manifold $M$: this defines a braided Lie bracket
\begin{align}\notag
[X,Y]_\frg^\star \ := \ [-,-]_\frg\big(\mathsf{F}^{-1}\triangleright(X\otimes Y)\big) \= \big[\mathsf{F}^{-1}_{(1)}\triangleright X\,,\,\mathsf{F}^{-1}_{(2)}\triangleright Y\big]_\frg \ ,
\end{align}
of $X,Y\in\frg$. Unlike the star-commutators which preserve the Lie algebra structure, by construction the braided Lie bracket is braided antisymmetric,
\begin{align}\notag
[X,Y]_\frg^\star \= -\big[\mathsf{R}_{\mathsf{F}(1)}^{-1}\triangleright Y\,,\,\mathsf{R}_{\mathsf{F}(2)}^{-1}\triangleright X\big]_\frg^\star \ ,
\end{align}
and obeys a braided Jacobi identity. 

Contrary to star-gauge symmetries, braided gauge symmetries define braided derivations: 
\begin{align}\notag
\delta_\xi(f\star g)\=\delta_\xi f\,\star\,g \, + \, \big(\mathsf{R}_{\mathsf{F}(1)}^{-1}\triangleright f\big)\,\star\,\delta_{\mathsf{R}_{\mathsf{F}(2)}^{-1}\triangleright\xi}\,g \ .
\end{align}
It should be stressed though that the notion of braided gauge symmetry is not so recent, as kinematical aspects of this idea appeared long before in e.g.~\cite{Brzezinski:1992hda}. What is new is the systematic organisation of the gauge invariant dynamics of these theories in the language of braided $L_\infty$-algebras as well as the computation of quantum correlation functions using the braided BV formalism and homological perturbation theory. In particular, the formalism is naturally tailored to handle noncommutative gravity with twisted diffeomorphism symmetry in the first order formalism~\cite{Ciric:2020eab,DimitrijevicCiric:2021jea,Szabo:2022edp}.
 
Further details about the UV/IR mixing problem in noncommutative scalar field theory and its resolution in braided quantum field theory can be found in the contribution of Marija Dimitrijevi\'c \'Ciri\'c to this volume. As with the Grosse-Wulkenhaar model, while technically appealing there is no explicit physical realisation yet of braided quantum field theories as effective theories, for instance in string theory or other models of quantum gravity, which typically dictate the use of the ordinary Feynman path integral measure.
 
\section{Yang-Mills Matrix Models and Fuzzy Field Theories}
\label{sec:matrixmodels}

To close off our overview of noncommutative field theories, in this final section we discuss the intimate and natural relationship between noncommutative Yang-Mills theory and matrix models, that collectively go under the name of `Yang-Mills matrix models'. We focus, in particular, on the natural ultraviolet regularizations that matrix models can provide through the finite size of matrices. Among other remarkable features which are absent in the commutative case, this can be used to provide a nonperturbative constructive definition of noncommutative gauge theory. These types of noncommutative field theories are also called fuzzy field theories, and they are a topic of vast investigation and activity in their own right, particularly for numerical investigations. Fuzzy field theories can be quantized in a completely rigorous way which avoids the functional analytic complications of continuum field theories. They are also interesting examples of noncommutative field theories.

\subsection{Reduced Models and Emergent Phenomena}
\label{sub:twistedreduced}

Consider the noncommutative Yang-Mills action \eqref{eq:NCYM} for a $U(1)$ gauge field $A_i(x)$ on Euclidean spacetime $\bbr^p$ with the Moyal-Weyl star-product \eqref{eq:MoyalWeylstar} and metric  $G=\unit$:
$$
S_{\text{\tiny YM}} \=-\frac1{4g_{\text{\tiny YM}} ^2}\,\int_{\bbr^p} \, \dd^px \ F_{ij}\star F^{ij} \ ,
$$
where the field strength tensor is given by
$$
F_{ij}\=\partial_i A_j-\partial_j
A_i-\ii[A_i\stackrel{\star}{,}A_j] \ . $$
This action is invariant under infinitesimal star-gauge transformations \eqref{eq:stargauge}. As discussed in~\cite{Lizzi:2001nd}, these generate an infinite-dimensional gauge symmetry group which can be identified with the infinite unitary group {$U(\infty)$}, similarly to the infinite unitary symmetry discussed in \S\S\ref{sub:GWmodel}. This owes to the fact that noncommutative gauge transformations generally mix colour and spacetime degrees of freedom through the star-product, and correspond to ``quantum deformations'' of canonical transformations of the Poisson manifold $(\bbr^p,\theta)$. This geometrical description is important for gravitational applications.

The appearance of $U(\infty)$ gauge symmetry can be best understood by rewriting the action in terms of the covariant coordinates introduced in \S\S\ref{sub:SWmaps}:
$$
X_i\=\theta_{ij}^{-1}\,x^j+A_i \ .
$$
Recall that in these coordinates, noncommutative gauge transformations \eqref{eq:stargauge} act as inner automorphisms of  the algebra $\CA(\bbr^p_\theta)$:
\begin{align}\nonumber
\delta_\lambda^\star\, X_i \= \ii[\lambda\stackrel{\star}{,}X_i] \ .
\end{align}
In particular, the derivative operators $\partial_i$ can be removed entirely from the noncommutative gauge theory action by representing them as the inner derivations $-\ii\theta^{-1}_{ij}\,[x^j\stackrel{\star}{,}-]$, so that
\begin{align}\nonumber
F_{ij} \= -\ii[X_i\stackrel{\star}{,}X_j] + \theta^{-1}_{ij} \ .
\end{align}

The covariant coordinates are elements of the abstract algebra  $\CA(\bbr^p_\theta)$. We use the Weyl transform of \S\S\ref{sub:GWmodel} to map $X_i$ to Hermitian elements \smash{$\widehat{X}_i$} of the algebra $$\mathcal{K}\=\mathcal{K}(\hil)$$ of compact operators acting on a separable Hilbert space $\hil$. In the following we drop the hat symbols on operators for brevity. The finite $U(\infty)$ gauge symmetry of noncommutative Yang-Mills theory is then realised through inner automorphisms parametrized by the gauge group $\CU(\CK)$:
\begin{align}\label{eq:gaugeIKKT}
X_i \ \longmapsto \ U\,X_i\,U^{\dag} \ , \quad U \ \in \ \CU(\CK) \ .
\end{align}
Informally, we may regard the covariant coordinates as infinite matrices $X_i\in\text{Mat}(\infty)$, where
\begin{align}\notag
\text{Mat}(\infty) \= \underset{N\in\mathbbm{N}}{\underrightarrow{\lim}} \, \text{Mat}(N)
\end{align} 
is an inductive limit of finite rank matrix algebras, which is dense in $\CK$ (with respect to the operator norm topology).

In this way spacetime derivatives completely disappear in the rewriting of noncommutative Yang-Mills theory in terms of covariant coordinates, and spacetime integration acts as a trace on the noncommutative algebra of fields $\CA(\bbr_\theta^p)$, represented on the Hilbert space $\hil$. Thus the noncommutative gauge theory becomes a {matrix model}~\cite{Aoki:1999vr,Madore:2000en}
\begin{align} \label{eq:IKKT}
S_{\text{\tiny YM}}\=-\frac1{4g_{\text{\tiny YM}}^2} \, \sum_{i, j=1}^p \,\Tr\,\big(-\ii[X_i,X_j]+\theta_{ij}^{-1}
\big)^2
\end{align}
with no reference to spacetime. The action \eqref{eq:IKKT} is manifestly invariant under unitary transformations \eqref{eq:gaugeIKKT}. This theory formally makes sense for finite rank matrices $X_i\in\mathfrak{u}(N)$, and even more generally when $X_i$ are elements of a quadratic Lie algebra $\frg$ with a central element. In particular, there are no non-trivial outer automorphisms, and all automorphisms of the noncommutative algebra of fields are realised as gauge symmetries in this rewriting.

We stress that although \emph{any} noncommutative field theory is formally a matrix model, insofar as any noncommutative algebra can be represented as operators on a separable Hilbert space, what is remarkable here is the disappearance of the infinitesimal outer automorphisms $\partial_i$ from the action. For instance, this is in marked contrast to the matrix model representation \eqref{eq:Scovmatrix} of the noncommutative scalar field theory \eqref{eq:Scov}, which even at the self-dual point \eqref{eq:selfdualpoint} involves an external field in its kinetic term. There is no analogue of this manipulation in ordinary Yang-Mills theory.

The large $N$ matrix model \eqref{eq:IKKT} is called a twisted reduced model. It can be regarded as the dimensional reduction of
  ordinary $U(N)$ Yang-Mills theory in $p$ dimensions to a point, i.e. by restricting the Yang-Mills action to constant gauge fields; the `twist'  $\theta_{ij}^{-1}$ is a background flux which modifies the vacuum structure of the theory. When $p=10$ the matrix model without the twist defines the bosonic part of the Ishibashi-Kawai-Kitazawa-Tsuchiya (IKKT) model, which conjecturally provides a concrete nonperturbative definition  of type~IIB string theory~\cite{Ishibashi:1996xs}. In particular, a natural regularization of the path integral is provided by the matrix rank $N$: the measure is the $N\to\infty$ limit of the translationally-invariant Haar measure inherited from the bi-invariant Haar measure on the  Lie group $U(N)^{\times p}$. The action \eqref{eq:IKKT} extends to a holomorphic function on the space $\bbc^p\otimes\text{Mat}(N)$.

The Yang-Mills equations of motion in this rewriting take the form
\begin{align} \label{eq:IKKTeom}
\sum_{i=1}^p\,[X_i,[X_i,X_j]]\=0 \ , \quad j\=1,\dots,p \ .
\end{align}
That is, the non-local partial differential equations for noncommutative Yang-Mills theory become \emph{algebraic} equations.
The vacuum solution of \eqref{eq:IKKTeom} is given by flat connections, $F_{ij}=0$, which satisfy
\begin{align}\label{eq:vacuum}
[X_i,X_j]\=-\ii\theta_{ij}^{-1} \ .
\end{align}
This shows that the noncommutative gauge degrees of freedom are fluctuations around the canonical noncommutative spacetime \eqref{eq:MoyalWeylcomm}. This is somewhat similar in spirit to the emergence of gauge fields through fluctuations of a Dirac operator that we discussed in~\S\S\ref{sub:StandardModel}.

However, this model is intrinsically infinite-dimensional and hence \emph{perturbative}: as usual with Heisenberg-type commutation relations, there are no finite $N\times N$ solutions to the vacuum equations \eqref{eq:vacuum}. But there is a finite-dimensional version of noncommutative gauge theory that provides a nonperturbative regularisation, and provides a constructive definition of noncommutative Yang-Mills theory as a quantum field theory within a rigorous framework.

A nonperturbative {finite {$N$}} definition of noncommutative Yang-Mills theory is provided by the action~\cite{Ambjorn:1999ts}
\begin{align} \label{eq:TEK}
S_{\text{\tiny TEK}}\=-\frac1{4g_{\text{\tiny YM}}^2} \ \sum_{1\leqslant i\neq j\leqslant p} \,\e^{-2\pi\ii Q_{ij}/N}~
\Tr\big(U_i\,U_j\,U_i^\dag\,U_j^\dag\big) \ ,
\end{align}
where $U_i\in U(N)$ and $Q_{ij}=-Q_{ji}\in\bbz$ for $i,j=1,\dots,p$. The path integral measure is the bi-invariant Haar measure on $U(N)^{\times p}$. By identifying the $N\times N$ unitary matrices $$U_i\=\exp(\ii a\,X_i) \ , $$ where $a$ is a dimensionful lattice spacing, in the double scaling limit $a\to0$ and $N\to\infty$ with  
$$\theta^{-1}_{ij}\=
  \frac{2\pi\,Q_{ij}}{N\,a^2}$$
held finite, the action \eqref{eq:TEK} becomes the action of the reduced model \eqref{eq:IKKT}.

The unitary matrix model defined by \eqref{eq:TEK} is the twisted Eguchi-Kawai model~\cite{Gonzalez-Arroyo:1982hyq}. It was originally derived as the one-plaquette reduction of Wilson's lattice gauge theory in $p$ dimensions with multivalued gauge fields and background 't~Hooft flux, which is equivalent to ordinary Yang-Mills gauge theory in the large $N$ planar 't~Hooft limit where $a\to0$ and $\theta_{ij}\to\infty$. The periodicity of the lattice in the present context is a nonperturbative form of UV/IR mixing; in particular, the noncommutative planar limit does not commute with the commutative limit $\theta_{ij}\to0$. 

For finite values of $N$, the matrix model \eqref{eq:TEK} corresponds to a noncommutative version of lattice gauge theory under a finite-dimensional version of the Weyl transform from \S\S\ref{sub:GWmodel}, where $N$ is related to the size of the periodic lattice  \cite{Ambjorn:2000nb,Ambjorn:2000cs}. The vacuum equations define the 't~Hooft algebra
\begin{align}\label{eq:tHooftalgebra}
U_i\,U_j\=\e^{2\pi\ii Q_{ij}/N} \ U_j\,U_i \ ,
\end{align}
which can be realised by twist-eating solutions constructed from $SU(N)$ clock and shift matrices. These are the defining relations of a $p$-dimensional `fuzzy torus', where one regards the generators $U_1,\dots,U_p$ as the $p$ exponential functions corresponding to the $p$ one-cycles of the torus. We will discuss this background further in~\S\S\ref{sub:fuzzytorus} for the case $p=2$.

Noncommutative tori also emerge as solutions in toroidal compactification of the IKKT matrix model~\cite{Connes:1997cr}. Let $\bbt^p$ be the rectangular $p$-dimensional torus whose $i$-th one-cycle has radius $R_i>0$. Compactification on $\mathbbm{T}^p$ is defined by restricting the action \eqref{eq:IKKT} to a subspace where an equivalence relation 
\begin{align}\label{eq:IKKTequiv}
 X_i \ \sim \ X_i+2\pi\,R_i\unit
\end{align} 
is satisfied by the covariant coordinates. The action is well-defined on this subspace if it depends only  on the equivalence classes, that is,
\begin{align} \nonumber
S_{\text{\tiny YM}}(X_i+2\pi\,R_i\unit) \= S_{\text{\tiny YM}}(X_i) \ , \quad i\=1,\dots,p \ .
\end{align}

By invariance of the IKKT model under finite unitary transformations of the covariant coordinates \eqref{eq:gaugeIKKT}, this can be achieved by realising the equivalence relation \eqref{eq:IKKTequiv} as unitary gauge equivalence through the \emph{quotient conditions}
\begin{align} \label{eq:quotientconds}
X_i+2\pi\,R_i\,\delta_{ij}\unit \= U_j\,X_i\,U_j^{-1} \ , \quad i,j\=1,\dots,p \ ,
\end{align} 
where $U_j$ are unitary. 
Taking the trace of both sides of \eqref{eq:quotientconds} shows that they cannot be solved by finite-dimensional matrices $X_i$ when $R_i\neq0$. Hence we seek perturbative solutions $X_i$ and $U_j$ which are operators on a separable infinite-dimensional Hilbert space~$\hil$.

It is straightforward to derive from the quotient conditions \eqref{eq:quotientconds} that, for each pair $i,j\in\{1,\dots,p\}$, the operator $U_i\,U_j\,U_i^{-1}\,U_j^{-1}$ commutes with all $X_k$ for $k=1,\dots,p$. The natural choice is to set it equal to a scalar operator $\lambda_{ij}\unit$ on $\hil$ with $\lambda_{ij}\in\bbc$. Unitarity implies that $\lambda_{ij}=\e^{2\pi\ii\theta_{ij}}$, for some $\theta_{ij}\in[0,1)$, which leads to the algebra
\begin{align} \nonumber
U_i\,U_j\=\e^{2\pi\ii\theta_{ij}} \ U_j\,U_i \ .
\end{align}
These are just the defining relations for the generators of the noncommutative torus $\bbt_\theta^p$. 

Now the quotient conditions \eqref{eq:quotientconds} are simply the Leibniz rule for the components $X_i$ of a connection on a projective module $\hil$ over the noncommutative torus $\bbt_\theta^p$. It follows that toroidal compactification of the IKKT matrix model is Yang-Mills theory on a noncommutative torus~\cite{Connes:1997cr}. The relation between the noncommutative torus $\bbt_\theta^p$ and the fuzzy torus defined by the 't~Hooft algebra \eqref{eq:tHooftalgebra} in the large $N$ limit is described in \cite{Landi:1999ey}; in particular, Morita equivalent noncommutative tori arise from the \emph{same} scaling limit of fuzzy tori in this sense.

More general noncommutative spacetimes are obtained as non-vacuum solutions of the Yang-Mills equations \eqref{eq:IKKTeom} satisfying
$$\big[X^i\,,\,X^j\big]\=\ii\theta^{ij}(X) \ .$$
Noncommutative geometries such as fuzzy spheres and other fuzzy spaces of various dimensions generically appear as classical solutions of Yang-Mills matrix models. They describe brane configurations in string theory, see e.g.~\cite{Alekseev:1999bs,Alekseev:2000fd} as well as~\cite{Szabo:2005jj} for a general review of how D-branes arise as states of noncommutative field theories. Thus (noncommutative) spacetime emerges as a dynamical effect in the matrix model.

This perspective has been used to develop models of emergent gravity, which clarifies the origin of gravity in noncommutative gauge theory~\cite{Rivelles:2002ez,Yang:2004vd,Steinacker:2007dq}. In $U(N)$ noncommutative Yang-Mills theory,  the {$U(1)$} ``photon'' is really a graviton, defining a
  non-trivial geometric background coupled to {$SU(N)$} gauge
  fields.  In this setting, a dynamical quantum spacetime relates gravity to quantum fluctuations of the covariant coordinates {$X^i$} of spacetime at the Planck scale,  and noncommutative field theory arises from field dependent fluctuations of spacetime geometry; again this is similar in spirit to the coupling of gravity to gauge fields through fluctuations of a Dirac operator discussed in \S\S\ref{sub:StandardModel}. This model is similar to general relativity for weak curvature, and it suggests a new
  approach to the quantization and unification of gravity with gauge
  theory; in particular, flat space is stable in this theory at one-loop order. See~\cite{Steinacker:2010rh} for an early review on the subject, while a more recent thorough account is found in the text~\cite{Steinacker:2024unq}.

Emergent gravity also provides a beautiful interpretation of the problem with UV/IR mixing discussed in \S\S\ref{sub:UVIRmixing}~\cite{Grosse:2008xr}. 
UV/IR mixing only arises for the abelian $U(1)\subset U(N)$ degrees of freedom.  
 Integrating out the noncommutative gauge fields involving the $U(1)$ sector UV/IR mixing terms in the region of momentum space $$
    k \ < \ \Lambda \ < \ \Lambda_{\text{\tiny NC}}\=\frac1{\sqrt{\theta}}$$ yields a one-loop
    effective action that includes an induced Einstein-Hilbert
    action, with the  gravitational constant {$G$} determined by the ultraviolet cutoff $\Lambda$. In this picture, UV/IR mixing arises due to a non-renormalizable gravitational sector in the infrared.
    
A somewhat different perspective on the emergence of (noncommutative) gravity in noncommutative gauge theory has also been recently established using more modern double copy techniques~\cite{Szabo:2023cmv,Szabo:2024dfa,Jonke:2025pkj}. See also~\cite{Langmann:2001yr} for an earlier perspective based on the teleparallel equivalent of general relativity.

In the remainder of this contribution we describe two explicit examples illustrating how effective field theories on the fuzzy space solutions of the Yang-Mills equations are built.
  
\subsection{Field Theory on the Fuzzy Sphere}

A classic example of a fuzzy space is the two-dimensional fuzzy sphere~\cite{Hoppe82,Madore:1991bw}, on which field theory can be defined and analysed using familiar techniques from the algebra of angular momentum in  quantum mechanics. For an integer $N>1$, consider the spin
$$\alpha\=\frac{N-1}2$$ irreducible representation $(\alpha)$ of the Lie algebra $\mathfrak{su}(2)$ of $SU(2)$, with dimension $N$. Its Hermitian generators $X_i=X_i^\dag$ for $i=1,2,3$ satisfy the fuzzy sphere relations which comprise the $\mathfrak{su}(2)$ Lie algebra
\begin{align}\label{eq:fuzzysphererels}
[X_i,X_j]\=\ii \ell_N\,\epsilon_{ijk}\,X_k \ ,
\end{align}
together with the Casimir constraint 
\begin{align}\label{eq:Casimir}
\sum_{i=1}^3 \, X_i^2\=R^2\unit  \ , 
\end{align}
where $R>0$ is the radius which is quantized in units of the noncommutativity parameter $\ell_N$ through $$\ell_N\=\frac{2R}{\sqrt{(N-1)\,(N+1)}} \ . $$ 
 
The quantization of $R$ is induced by its relation  \eqref{eq:Casimir} to the quadratic Casimir invariant of the spin~$\alpha$ representation of $\mathfrak{su}(2)$, which is quantized. The generators $X_i$ span the unital $*$-algebra
  $$\CA(\mathbbm{S}_N^2)\=\text{End}_\bbc(\alpha) \= (\alpha)\otimes(\alpha)^*\ \simeq \ \text{Mat}(N) \ , $$
which we regard as the `algebra of functions' on a fuzzy sphere $\mathbbm{S}_N^2$ of radius $R$. Under this isomorphism we identify the $\mathfrak{su}(2)$ Lie bracket with the matrix commutator. As an $\mathfrak{su}(2)$-module, the algebra $\CA(\mathbbm{S}_N^2)$ decomposes into irreducible representations as
\begin{align}\label{eq:fuzzyS2irreps}
\CA(\mathbbm{S}_N^2) \ \simeq \ \bigoplus_{J=0}^{N-1} \, (J) \ .
\end{align}

Differential operators representing the components $L_i$ of angular momentum on the fuzzy sphere are realised as the inner derivations $\frac1{\ell_N}\,[X_i,-]$ of the algebra $\CA(\mathbbm{S}_N^2)$. 
Free real scalar field theory on the fuzzy sphere is then defined by the action~\cite{Grosse:1995ar}   $$ S_0\=\frac12\,\frac{4\pi}N\Tr\,\Phi\,\big(\Delta+m^2\big)\,\Phi$$ 
for a Hermitian element $\Phi\in\CA(\mathbbm{S}_N^2)$, where the fuzzy Laplacian $\Delta:\CA(\mathbbm{S}_N^2)\longrightarrow\CA(\mathbbm{S}_N^2)$ is the linear map
$$\Delta(\Phi)\=\frac1{\ell_N^2}\,\sum_{i=1}^3\,[X_i,[X_i,\Phi]] \ . $$ 

The  fuzzy Laplacian is diagonalised by the $N^2$ fuzzy spherical harmonics $$Y_j^J \ \in \ \CA(\mathbbm{S}_N^2)$$ for $0\leqslant  J\leqslant  N-1$ and $-J\leqslant  j\leqslant  J$, with the eigenvalue equation $$\Delta(Y_j^J)\=J\,(J+1)\,Y_j^J \ . $$
They form an orthonormal basis of $\CA(\mathbbm{S}_N^2)$,
$$\frac{4\pi}N \Tr(Y_j^J{}^\dag\,Y_{j'}^{J'})\=\delta_{JJ'}\,\delta_{jj'} \ , $$
satisfy the reality condition
\begin{align}\nonumber
Y_j^J{}^\dag \= (-1)^J\,Y^J_{-j} \ ,
\end{align}
and obey the associative fusion algebra
\begin{align*}
Y_i^I \, Y_j^J & \= \displaystyle \sum_{\stackrel{\scriptstyle 0\leqslant K\leqslant N-1}{\scriptstyle -K\leqslant k\leqslant K}} \, (-1)^{2\alpha+I+J+K+k} \, \sqrt{(2I+1)\,(2J+1)\,(2K+1)} \\[-1em]
& \hspace{3cm} \times \, \left({ \begin{matrix} I & J & K \\ i & j & -k \end{matrix} }  \right) \, \left\{ {\begin{matrix} I & J & K \\ \alpha & \alpha & \alpha \end{matrix} } \right\} \ Y_k^K \ ,
\end{align*}
where $(\quad)$ (resp. $\{\quad\}$) are the Wigner $3j$ (resp. $6j$) symbols of $\mathfrak{su}(2)$.

When the scalar fields $\Phi$  are expanded in the basis of fuzzy spherical harmonics, interaction vertices in this theory are expressed in terms of $3j$ and $6j$ symbols. For the interacting action of $\Phi^4$-theory
\begin{align}\nonumber
S_{\rm int} \= \frac\lambda{4!} \, \frac{4\pi}N \Tr\,\Phi^4 \ , 
\end{align}
the vertex is given by
\begin{align}\nonumber
I_{j_1\cdots j_4}^{J_1\cdots J_4} & \=  \frac{\lambda}{4!} \, \frac N{4\pi} \, (-1)^{J_1+\cdots + J_4} \, \prod\limits_{i=1}^4\, \sqrt{2J_i+1} \ \sum\limits_{\stackrel{\scriptstyle 0\leqslant J\leqslant N-1}{\scriptstyle -J\leqslant j\leqslant J}}\, (-1)^j\,(2J+1) \\ 
& \hspace{1cm} \, \times\,\left({ \begin{matrix} J_1 & J_2 & J \\ j_1 & j_2 & j \end{matrix} }  \right) \, \left({ \begin{matrix} J_3 & J_4 & J \\ j_3 & j_4 & -j \end{matrix} }  \right)\,\left\{{\begin{matrix} J_1 & J_2 & J \\ \alpha & \alpha & \alpha \end{matrix} }  \right\} \, \left\{{ \begin{matrix} J_3 & J_4 & J \\ \alpha & \alpha & \alpha \end{matrix}}  \right\}  \ . \label{eq:fuzzyspherephi4}
\end{align}
This vertex is invariant under cyclic permutations of the pairs $(J_i,j_i)$ for $i=1,\dots,4$.

To quantise the theory with action $S_0+S_{\rm int}$, we follow the path integral quantization prescription proposed originally in~\cite{Grosse:1995ar}. The path integral is defined by the finite-dimensional integral over the fuzzy angular momentum modes of the scalar fields. For example, the propagator is
\begin{align}\label{eq:fuzzysphereprop}
\varDelta^{JJ'}_{jj'} \= \delta_{JJ'} \, \delta_{jj'} \ \frac1{J\,(J+1)+m^2} \ .
\end{align}
The one-particle irreducible two-point function at one-loop order is obtained by contracting two legs in \eqref{eq:fuzzyspherephi4} using the propagator \eqref{eq:fuzzysphereprop}. This leads to the finite sum
\begin{align}\label{eq:fuzzyS22pt}
\varGamma^{(2)KK'}_{kk'} \= \frac\lambda{96\pi} \, \delta_{KK'} \, \delta_{k+k',0} \ \sum\limits_{\stackrel{\scriptstyle 0\leqslant J\leqslant N-1}{\scriptstyle -J\leqslant j\leqslant J}}\, \frac{I^{KJJK'}_{kjjk'}}{J\,(J+1)+m^2} \ .
\end{align}

The two-point function \eqref{eq:fuzzyS22pt} receives both planar and non-planar contributions~\cite{Chu:2001xi}, corresponding to whether neighbouring or non-neighbouring legs are contracted. The commutative limit is $N\to\infty$ with $R$ fixed, whereby the commutators in \eqref{eq:fuzzysphererels} vanish and the algebra $\CA(\mathbbm{S}_N^2)$ becomes the commutative algebra of functions on the usual sphere $\mathbbm{S}^2$ of radius $R$; see~\cite{Rieffel:2001qw} for a precise rigorous description of this limit. In this limit the planar contribution coincides precisely with the corresponding logarithmically divergent terms on the classical sphere, whereas the non-planar contribution is an analytic function of the noncommutativity parameter $\frac1\alpha$; that is, no infrared singularity develops and there is no UV/IR mixing problem on the fuzzy sphere. On the other hand, in the double scaling limit $N\to\infty$ and $R\to\infty$ with $$\theta\=\frac{2R^2}N$$ fixed, the fuzzy sphere $\mathbbm{S}_N^2$ becomes the Moyal plane $\bbr_\theta^2$, and one recovers the UV/IR mixing phenomenon~\cite{Chu:2001xi}. 

By construction, the fuzzy sphere algebra $\CA(\mathbbm{S}_N^2)=(\alpha)\otimes(\alpha)^*$ is covariant under the adjoint action of the Lie algebra $\mathfrak{su}(2)$. It has a well-known three-dimensional $\mathfrak{su}(2)$-equivariant differential calculus $\Omega^\bullet\CA(\mathbbm{S}_N^2)$, the Chevalley-Eilenberg differential graded algebra of $\mathfrak{su}(2)$ in the representation \eqref{eq:fuzzyS2irreps}:
\begin{align}\notag
\Omega^\bullet\CA(\mathbbm{S}_N^2) \= \CA(\mathbbm{S}_N^2)\otimes\midwedge^\bullet \mathfrak{su}(2)^* \ .
\end{align} 
This enables the construction of fuzzy field theories with gauge symmetries, such as Chern-Simons and Yang-Mills theories, among others; see e.g.~\cite{Madore:1991bw,Klimcik:1997mg,Grosse:1998gn,Carow-Watamura:1998zks} for early works on the subject. The action with Yang-Mills and Chern-Simons terms describes the low-energy effective field theory on D-branes in the $SU(2)$ WZW model, viewed as the target space $\mathbbm{S}^3$ with $H$-flux at large radius~\cite{Alekseev:1999bs}.
    
\subsection{Field Theory on the Fuzzy Torus}
\label{sub:fuzzytorus}

The two-dimensional fuzzy torus is another classic example of a fuzzy space which dates back to Weyl's early work on quantum mechanics~\cite{Weyl31}. It also serves as an interesting and tractable finite-dimensional example for structural constructions in noncommutative geometry~\cite{Landi:1999ey,Barnes:2016bmg,Barrett:2019ize}. For an integer $N>1$, the unital $*$-algebra $\CA(\bbt_N^2)$ of the fuzzy torus $\bbt_N^2$ is the 't~Hooft algebra $\bbc[U_1,U_2]$ whose unitary generators $U_i$ for $i=1,2$ are subject to the fuzzy torus relations
\begin{align*}
U_1\,U_2&\=q\,U_2\,U_1 \ , \\[4pt]
 U_i^N&\=\unit \ ,
 \end{align*}
where $$q\=\e^{2\pi\ii/N}$$ is a primitive $N$-th root of unity. 
The algebra $\CA(\bbt_N^2)$ is also isomorphic to the matrix algebra $$\CA(\bbt_N^2) \ \simeq \ \text{Mat}(N) \ , $$ which can be realised explicitly by representing $U_1$ and $U_2$ respectively as the traceless $SU(N)$ clock and shift matrices
\begin{align}\nonumber
U_1&\={\small\begin{pmatrix}
1 & 0 & \cdots & 0 \\
0 & q & \cdots & 0 \\[0.01mm]
0 & 0 & \ddots & 0 \\
0 & 0 & 0 & q^{N-1}
\end{pmatrix}\normalsize} \ , \\[4pt] 
U_2&\={\small\begin{pmatrix}
0 & 0 & \cdots & 1 \\
1 & 0 & \cdots & 0 \\[0.01mm]
0 & \ddots & 0 & 0 \\
0 & 0 & 1 & 0
\end{pmatrix}\normalsize} \ . \notag
\end{align}

As previously, free scalar field theory on the fuzzy torus is defined by the action $$ S_0\=\frac12\,\frac1N\Tr\,\Phi\,\big(\Delta+m^2\big)\,\Phi$$ 
for a Hermitian element $\Phi\in\CA(\bbt_N^2)$, where the fuzzy Laplacian $\Delta:\CA(\bbt_N^2)\longrightarrow\CA(\bbt_N^2)$ is the linear map
$$ \Delta(\Phi)\=-\frac{1}{\big(q^{1/2} - q^{-1/2}\big)^2}\,
\sum_{i=1}^2\,\big[U_i,\big[U_i^\dag,\Phi\big]\big] \ . $$
It is diagonalised by the $N^2$ fuzzy plane waves $$e_{{\underline k}}\= U_1^{k_1}\,U_2^{k_2} \ \in \ \CA(\bbt_N^2) \ , $$ for $\underline k=(k_1,k_2)\in\bbz_N^{\times 2}$, where $\bbz_N=\{0,1,\dots,N-1\}$ is the cyclic group of order $N$. The eigenvalue equation is $$\Delta(e_{{\underline k}})\=\big([k_1]^2_q + [k_2]^2_q\big)\, e_{{\underline k}} \ , $$ where $$[n]_q \= \frac{q^{n/2} - q^{-n/2}}{q^{1/2}-q^{-1/2}}\=\frac{\sin\big(\frac{\pi\,n}N\big)}{\sin\big(\frac\pi N\big)}$$ for $n\in\bbz$ are $q$-numbers. 

The fuzzy plane waves form an orthonormal basis
$$\frac1N \Tr(e_{{\underline k}}^\dag\, e_{{\underline l}}\,) \= \delta_{{\underline k},{\underline l}} \ . $$
They also satisfy the reality property
\begin{align}\nonumber
e_{\underline k}^\dag \= q^{-k_1\,k_2} \ e_{-\underline k} \ ,
\end{align}
and obey the fusion algebra
\begin{align}\nonumber
e_{\underline k} \, e_{\underline l} \= q^{-l_1\,k_2} \ e_{\underline k + \underline l} \ .
\end{align}

Consider now the interacting action of $\Phi^4$-theory
\begin{align}\nonumber
S_{\rm int} \= \frac\lambda{4!} \, \frac{1}N \Tr\,\Phi^4 \ , 
\end{align}
with the scalar fields $\Phi$ expanded in the basis of fuzzy plane waves. This leads to the interaction vertex
$$I_{\underline{k}{}_1\cdots\underline{k}{}_4}\= \frac\lambda{4!} \ q^{\,\sum\limits_{i<j}\, {k_i}{}_1\, {k_j}{}_2} \ \delta_{\underline{k}{}_1+\cdots +\underline{k}{}_4,\underline{0}} \ . $$
Again this vertex is invariant under cyclic permutations of $\underline{k}{}_i$ for $i=1,\dots,4$.

Path integral quantization is defined by the finite-dimensional integral over the fuzzy Fourier modes of the scalar fields. Then the propagtor is 
\begin{align}\nonumber
\varDelta_{\underline{k},\underline{l}} \= \delta_{\underline{k},\underline{l}} \ \frac1{[k_1]^2_q + [k_2]^2_q + m^2} \ .
\end{align}
The one-particle irreducible two-point function at one-loop order is given by the finite sum
\begin{align}\nonumber
\varGamma^{(2)}_{\underline{p},\underline{p}'} \= \frac\lambda{24} \, \delta_{\underline{p}+\underline{p}',\underline{0}} \ \sum_{\underline{k}\in\bbz_N^{\times 2}} \, \frac{I_{\underline{p}\,\underline{k}\,\underline{k}\,\underline{p}'}}{[k_1]^2_q + [k_2]^2_q + m^2} \ ,
\end{align}
receiving again contributions from both planar and non-planar Feynman diagrams. Scaling limits on the fuzzy torus as $N\to\infty$ are discussed in~\cite{Landi:1999ey} and related to continuum field theories on both commutative as well as noncommutative tori.

The triangular group Hopf algebra $$\mathbbmss{H}\=\bbc[\bbz_N^{\times 2}]$$ 
has structure maps given by
\begin{align}\notag
\mbf\Delta(\underline{k})&\=\underline{0}\otimes\underline{k} + \underline{k}\otimes\underline{0} \ , \\[4pt]
 \mathsf{S}(\underline{k})&\=-\underline{k} \notag \ , \\[4pt] \varepsilon(\underline{k})&\=0 \ , \notag
\end{align}
for $\underline{k}\in\bbz_N^{\times 2}$, and extended linearly. It acts on $\CA(\bbt_N^2)$ via
$$\underline{k}\triangleright U_i \= q^{k_i}\, U_i \ , \quad i\=1,2 \ . $$
This makes $\CA(\bbt_N^2)$ into a braided commutative $\mathbbmss{H}$-module algebra, with R-matrix
\begin{align}\nonumber
\mathsf{R}\=\frac1{N^2} \, \sum_{\underline s,\underline t\in\bbz_N^{\times 2}} \, q^{s_2\,t_1-s_1\,t_2} \ \underline s\otimes \underline t \ \in \ \mathbbmss{H}\otimes\mathbbmss{H} \ .
\end{align}
Thus scalar field theory on the fuzzy torus can also be quantized as a  {braided} quantum field theory~\cite{Nguyen:2021rsa}. However, what does not seem to be presently available is a suitable differential calculus on the fuzzy torus that enables extensions to (ordinary or braided) gauge theories.


\begin{ack}
The author would like to gratefully thank the organizers of the conference, Roberta Iseppi, Pierre Martinetti, Thierry Masson, Gaston Nieuviarts and Patrizia Vitale, for the invitation to deliver this mini-course and for seamlessly putting together a highly enjoyable conference. Special thanks are due as well to Joseph Kouneiher for his excellent organisation of this Special Volume.
\end{ack}

\begin{funding}
This article is based upon work from COST Actions CaLISTA CA21109 and THEORY-CHALLENGES CA22113 supported by COST (European Cooperation in Science and Technology).
\end{funding}


\bibliographystyle{emss}  
\bibliography{Szabo-Marseille.bib}

\end{document}